\documentclass{article}
\usepackage[utf8]{inputenc}
\usepackage{graphicx}
\usepackage{tabularx}
\usepackage{fancyhdr}
\usepackage{indentfirst}
\usepackage{epsf}
\usepackage{latexsym,euscript}
\usepackage{amsmath}
\usepackage{amsfonts}
\usepackage{amssymb, epsfig}
\usepackage{subcaption}
\usepackage[]{units}
\usepackage{multirow}
\usepackage{verbatim}
\usepackage{geometry}
\usepackage{longtable}
\usepackage{booktabs}
\usepackage{float}
\usepackage{tikz}
\usepackage{dsfont}
\usepackage{braket}
\usepackage{blkarray}
\usepackage{comment}
\usepackage{lscape}
\usetikzlibrary{quantikz}
\usepackage[pdftex, pdfstartview={FitH}, pdfnewwindow=true, colorlinks=true, citecolor=blue, filecolor=blue, linkcolor=blue, urlcolor=blue, pdfpagemode=UseNone]{hyperref}
\usepackage{cleveref}
\usepackage[titletoc]{appendix}
\usepackage{enumitem}
\usepackage{soul}
\usepackage{array}
\newcolumntype{L}[1]{>{\raggedright\arraybackslash}p{#1}}

\usepackage[english]{babel}
\usepackage[babel]{csquotes}
\usepackage[natbib, maxnames=999, minnames=1, maxcitenames=2, mincitenames=1, maxbibnames=999, minbibnames=1, bibstyle=numeric, citestyle=numeric-comp, backend=bibtex, backref, hyperref, sorting=none, giveninits=true]{biblatex}

\usepackage{tabularray}
\usepackage[export]{adjustbox}
\usepackage{microtype}
\usepackage[T1]{fontenc}

\bibliography{SQM__Bib.bib}
\renewbibmacro{in:}{}

\newcommand{\qhat}{\ensuremath{\hat{q}} }
\newcommand{\phat}{\ensuremath{\hat{p}} }

\newcommand{\bhat}{\ensuremath{\hat{b}} }
\newcommand{\bhatdag}{\ensuremath{\hat{b}^\dag} }

\newcommand{\La}{\ensuremath{\Lambda} }
\newcommand{\eq}[1]{Eq.~(\ref{#1})}
\newcommand{\fig}[1]{Fig.~\ref{#1}}
\newcommand{\tab}[1]{Table~\ref{#1}}

\begin{document}
\begin{flushright}
\today\\
\end{flushright}

\vspace{0.1cm}

\begin{center}
  {\Large \textbf{Quantum Variational Methods \\ for Supersymmetric Quantum Mechanics}}
\end{center}

\vskip 0.3cm
\centerline{John Kerfoot, Emanuele Mendicelli, David Schaich}

\vskip 0.6cm

\centerline{\textsl{Department of Mathematical Sciences, University of Liverpool,}}
\centerline{\textsl{Liverpool L69 7ZL, United Kingdom}}

\vskip 1 cm

\begin{abstract}
We employ quantum variational methods to investigate a single-site interacting fermion--boson system---an example of a minimal supersymmetric model that can exhibit spontaneous supersymmetry breaking. Our study addresses the challenges inherent in calculating mixed fermion--boson systems and explores the potential of quantum computing to advance their analysis. By using adaptive variational techniques, we identify optimal ans\"atze that scale efficiently, allowing for reliable identification of spontaneous supersymmetry breaking. This work lays a foundation for future quantum computing investigations of more complex and physically rich fermion--boson quantum field theories in higher dimensions.
\end{abstract}

\vskip 1 cm

\vfill
\hrule
\vspace{0.3cm}

\noindent \textit{e-mail addresses}: john.kerfoot@liverpool.ac.uk, e.mendicelli@liverpool.ac.uk, david.schaich@liverpool.ac.uk

\newpage

\tableofcontents

\section{Introduction and motivations}
\label{sec:introduction}

Quantum computing holds immense promise for advancing our understanding of nature by increasing our ability to simulate quantum systems for which classical computational methods suffer from exponentially growing costs.
In the last few years there has been extensive exploration of the capabilities of available `noisy intermediate-scale quantum' (NISQ~\cite{Preskill:2018jim}) devices and their usefulness for particle physics; recent reviews include Refs.~\cite{Bauer:2022hpo, DiMeglio:2023nsa, Funcke:2023jbq, Davoudi:2025kxb}.
Quantum variational methods in particular have received much attention, largely in the context of state preparation, which is necessary to initialize simulations of complex dynamical processes such as particle collisions~\cite{Farrell:2024fit, Farrell:2025nkx}.
Variational quantum algorithms have also shown significant promise beyond fundamental physics, in addressing challenging eigenvalue problems that arise in highly entangled quantum systems, especially those outside equilibrium~\cite{Halimeh:2025vvp}.
Applications in computational biology and quantum chemistry \cite{RevModPhys.92.015003} include modeling protein folding dynamics \cite{Robert_2021}, understanding reaction mechanisms, and predicting the structure and behavior of complex molecules \cite{Sivakumar:2024cia}.

In this work we apply quantum variational methods to analyze simple supersymmetric systems that suffer from sign problems in classical lattice quantum field theory approaches.
Supersymmetry extends the Poincaré spacetime symmetry group by adding fermionic generators that relate bosonic and fermionic fields.
This increased symmetry offers a vast array of applications, ranging from hypothetical extensions of the Standard Model~\cite{Allanach:2024suz}, to deepening our understanding of the foundational structures of quantum field theory~\cite{Weinberg:2000cr}, and through holographic dualities further advancing the exploration of quantum gravity~\cite{Maldacena:1997re}.
Despite the potential role of supersymmetry in protecting electroweak symmetry breaking from large quantum corrections, the persistent non-observation of superpartners of Standard Model particles implies that any viable supersymmetric description of the universe must include spontaneous supersymmetry breaking. This mechanism allows superparticles to be heavier than their Standard Model counterparts, thereby remaining consistent with current experimental limits.
While the study of spontaneous supersymmetry breaking is therefore of great interest, it is challenging as a generically non-perturbative phenomenon. 

The standard approach to studying non-perturbative phenomena is through lattice regularization of the theory, followed by numerical analyses using Monte Carlo methods. Despite significant progress in the lattice study of various non-perturbative aspects of supersymmetric quantum field theories~\cite{Kadoh:2016eju, Bergner:2016sbv, Schaich:2022xgy}, spontaneous supersymmetry breaking remains largely inaccessible.
This is because it requires a vanishing Witten index \cite{Witten:1982df}, which corresponds to an exactly zero partition function in in the context of lattice field theory.
This implies a severe sign problem, rendering standard Monte Carlo numerical calculations extremely challenging.\footnote{While in certain cases the sign problem can be avoided by reformulating the supersymmetric lattice system~\cite{Steinhauer:2014yaa}, this is not generally applicable.}

One possible way to circumvent the sign problem is to utilize quantum computers, which can efficiently access exponentially large Hilbert spaces, making the study of the theory in the Hamiltonian formalism -- where no sign problem is present -- conceptually viable.
Although there have been significant advancements in quantum computing over the past decade, current hardware remains in the NISQ stage. These devices are characterized by a limited number of qubits, noisy gate operations, and the absence of quantum error correction, which makes them challenging to apply to problems that are not already classically tractable.

While we await the availability of fault-tolerant quantum computers, it is important to explore and develop quantum computational methods for small-scale systems such as the supersymmetric lattice theories we consider here. These explorations can productively consider small lattice sizes in a low number of space-time dimensions, which remain accessible to exact diagonalization using classical computation. This enables the validation of studies on current quantum hardware, helping to develop quantum algorithms and best practices for future calculations using larger-scale and fault-tolerant quantum devices. In particular, supersymmetric systems provide an ideal sandbox to explore the challenges of encoding boson--fermion systems on current hardware.

There has been limited consideration of quantum computation for supersymmetric systems in recent years, including discussion of their classical and quantum computation complexity~\cite{Crichigno:2020vue, Cade:2021jhc}.
In Ref.~\cite{Cai:2022yup}, a spin--phonon coupling in a trapped-ion quantum simulator was identified with a particular supersymmetric Hamiltonian.
Finally, a series of conference proceedings~\cite{Culver:2021rxo, Culver:2023iif, Schaich:2024bmg, Mendicelli:2024ryt} explored the application of variational methods to identify spontaneous supersymmetry breaking in supersymmetric quantum mechanics (SQM) and the $\mathcal N = 1$ Wess--Zumino model.
In this work, we build on Refs.~\cite{Culver:2021rxo, Culver:2023iif, Schaich:2024bmg, Mendicelli:2024ryt}, focusing on variational quantum analyses of minimal SQM in 0+1~dimensions --- i.e., interacting fermion--boson systems on a single lattice site in.
In particular, we extend the earlier work by exploring the role of shot noise in these investigations.
Another key new component of this work is the use of adaptive methods to construct variational ans\"atze that scale efficiently as the number of qubits increases.
This provides a foundation for practical calculations using currently available quantum hardware, as well as extensions to more complex interacting fermion--boson systems such as supersymmetric matrix models and higher-dimensional quantum field theories, where a key challenge will be to maintain the scalability of quantum resources required.

This paper is organized as follows. We begin in Section~\ref{sec:susy_qm} with a detailed discussion of the SQM Hamiltonian, including the connection between spontaneous supersymmetry breaking and the ground-state energy.
This is followed in Section~\ref{sec:regularization} by an analysis of its regularization and the effects of truncating the bosonic degree of freedom. The qubit digitization is presented in Section~\ref{sec:fock_basis_digitization}.
In Section~\ref{sec:variational_methods}, we explore quantum variational methods for finding the ground state, starting with a general overview and emphasizing the connections between the initial state, the ansatz, and the classical optimizer. We then focus on the application of the Variational Quantum Eigensolver (VQE) using both local optimizers, such as COBYLA, and global ones like Differential Evolution.
As mentioned above, we have developed adaptive methods to identify more efficient ans\"atze for our work.
In Section~\ref{section:Adaptive-VQE} we present our method, while Section~\ref{ansatz_extrapolation} demonstrates the extrapolation of the resulting ans\"atze to an arbitrary number of qubits.
Finally, we can also exploit low-lying excited states to probe spontaneous supersymmetry breaking in a way that may be less sensitive to noise, which we discuss in Section~\ref{sec:vqd}, presenting results obtained using the Variational Quantum Deflation (VQD) algorithm.
We conclude in Section~\ref{sec:conclusions_f_directions} by summarizing our main findings and directions for future research, particularly in tackling more computationally demanding and physically rich supersymmetric systems, such as the Wess--Zumino model.

\section{Supersymmetric quantum mechanics}
\label{sec:susy_qm}

\begin{figure}[htbp]
\centering
\includegraphics[width=0.2\textwidth]{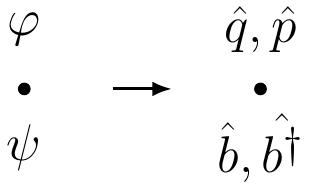}
  \caption{An illustration of the system, where the fermionic degree of freedom is represented by $\psi$ and the bosonic degree of freedom by $\phi$. In \eq{eq:H_SQM} these correspond to the operators $\qhat, \phat$ for the boson, and $\bhat, \bhatdag$ for the fermion.}
\label{fig:single_site}
\end{figure}

The SQM theory we consider is a discrete Hamiltonian model with one fermionic degree of freedom and one bosonic degree of freedom interacting at a single site in continuous time, as illustrated by \fig{fig:single_site}:
\begin{equation} \label{eq:H_SQM}
H = \frac{1}{2}\left( \phat^2 + [W'(\qhat)]^2 - W''(\qhat) \left[\bhatdag, \bhat\right]\right),
\end{equation}
where \phat and \qhat represent the momentum and position coordinate operators of the boson, with standard relations $[\qhat,\phat]=i$.
The interactions between bosonic and fermionic degrees of freedom is governed by the superpotential $W(\qhat)$, where the prime symbol indicates differentiation with respect to $\qhat$.
The operators \bhat and \bhatdag represent the fermionic annihilation and creation operators, and satisfy the standard canonical anticommutation relation $\lbrace \bhat, \bhatdag \rbrace = \mathds{1}$.
Their action on fermion states is:
\begin{align}
  \bhat \vert 1 \rangle & = \vert 0 \rangle   & \bhatdag \vert 1 \rangle & = 0 \cr
  \bhat \vert 0 \rangle & = 0           & \bhatdag \vert 0 \rangle & = \vert 1 \rangle.
  \label{b_b_dag_on_states}
\end{align}

The superpotential is crucial, as it governs whether supersymmetry is spontaneously broken or preserved~\cite{Cooper:1994eh, Woit:2017vqo}.
We consider three specific superpotentials in this work:
\begin{center}
\addtolength{\tabcolsep}{12 pt} 
\begin{tabular}{ccc}
  Harmonic Oscillator (HO) & Anharmonic Oscillator (AHO) & Double Well (DW) \\[5pt]
  $W(\qhat)=\frac{1}{2}m\qhat^2$ & $W(\qhat)=\frac{1}{2}m\qhat^2 + \frac{1}{4} g \qhat^4$ & $W(\qhat)=\frac{1}{2}m\qhat^2 +g \left( \frac{1}{3}\qhat^3 + \mu^2 \qhat  \right)$ \\[5pt]
  [Supersymmetric] & [Supersymmetric] & [Spontaneously broken]
\end{tabular}
\addtolength{\tabcolsep}{-12 pt} 
\end{center}
Here $m$ is the dimensionless mass of both the boson and fermion, while $g$ and $\mu$ are dimensionless interaction strengths.
In the body of the paper we set $m=g=\mu=1$ for simplicity, while Appendix~\ref{app:energyspectrum_artifacts} collects some representative results obtained for different values of these parameters.

As mentioned in Section~\ref{sec:introduction}, the supersymmetry of the system is generated by fermionic supercharges, which are also responsible for establishing the relation between the bosonic and fermionic degrees of freedom:
\begin{align}
  Q & = \bhat \left( i \phat + W'(\qhat) \right) &
  Q^{\dag} & = \bhatdag \left( -i \phat + W'(\qhat) \right) &
  Q^2 & = (Q^{\dag})^2 = 0.
\end{align}
The supercharges can be used to express the Hamiltonian as $2 H=\lbrace Q, Q^{\dag} \rbrace$.
This characteristic supersymmetric relation has profound effects on the energy spectrum.
First, all eigenstates have non-negative energy, given by $E_{\Psi} = \bra{\Psi} H \ket{\Psi} =\frac{1}{2}\left( \vert Q \ket{\Psi} \vert ^2  + \vert Q^{\dag} \ket{\Psi} \vert ^2\right) \geq 0$.
Second, an eigenstate has $E_\Psi=0$ only if it is annihilated by both supercharges, $(Q^{\dag} + Q) \ket{\Phi} = 0$.
In other words, a vanishing ground-state energy corresponds to the preservation of supersymmetry.
Finally, eigenstates with positive energy $E_{\Psi} > 0$ appear in degenerate pairs, which are related by the action of the supercharges.
The case of spontaneous supersymmetry breaking, where the supercharges do not annihilate the ground state, therefore corresponds to a degenerate pair of ground states with non-zero energy.

Thus the preservation or spontaneous breaking of supersymmetry for a given superpotential can be unequivocally determined by measuring the ground-state energy of the system.
This result holds for any supersymmetric theory, not just (0+1)-dimensional SQM.
Although this procedure may appear straightforward, it becomes computationally intensive and expensive for systems with a large number of states, as encoding the full Hilbert space requires an exponentially increasing amount of classical resources.
In this regard, quantum computing presents a promising alternative, by providing in principle a more efficient means of analyzing the Hilbert space and constructing the ground state.

As an aside, while the relation between spontaneous supersymmetry breaking and the ground-state energy is generic, the special case of SQM with a polynomial superpotential is simple enough that the preservation or spontaneous breaking of supersymmetry can be read off directly from the superpotential~\cite{Woit:2017vqo, Gendenshtein:1985hgo}.
Specifically, if the polynomial degree of the superpotential is even, then a normalizable solution to $(Q^{\dag} + Q) \ket{\Phi} =0$ can be found, implying preserved supersymmetry.
This is not the case for systems with an odd-degree superpotential, which therefore exhibit spontaneous supersymmetry breaking.
These expectations are noted in brackets under each superpotential listed above.

\section{Regularization of the model}
\label{sec:regularization}

To enable the encoding of the theory on classical or quantum hardware with finite resources, the infinite-dimensional Hilbert space must be made finite and manageable.
This is achieved by introducing a cutoff on the accessible states that makes them finite in a controlled way, a process known as regularization.

In our specific case of SQM, the infinite dimensionality of the Hilbert space arises from the bosonic degree of freedom, while the two fermionic states pose no issue.
To regularize the system, it is therefore sufficient to restrict the allowed bosonic modes.
We do so by truncating the possible excitation levels the bosonic degree of freedom can reach, in the Fock basis, retaining the lowest $\La = 2^B$ states, where $B$ is the number of qubits used to represent the bosonic degree of freedom.
Full details are in Section~\ref{sec:fock_basis_digitization}.

While regularization is essential for rendering the theory computationally tractable, it will introduce unphysical artifacts, which we examine in more detail in Appendix~\ref{app:energyspectrum_artifacts}.
In particular, in the context of supersymmetric systems, the different treatment of bosonic and fermionic degrees of freedom explicitly breaks supersymmetry.
These effects are expected to be mild except in situations where the system accesses the highest allowed excited states and becomes sensitive to the finiteness of the truncated Hilbert space.
The artifacts are addressed by removing the truncation $\La \to \infty$, systematically increasing the number of allowed bosonic modes.
This concretely means that systems with a range of \La must be studied to conclusively establish the spontaneous breaking or preservation of supersymmetry.

The regularized Hilbert space of the system consists of all possible combinations of fermionic and bosonic states
\begin{equation}
  \mathcal{H} = \mathrm{Span}\{\ket{f} \otimes \ket{b} \},
\end{equation}
where the fermionic degree of freedom $\ket{f} \in \left\{\ket{0}, \ket{1}\right\}$ and the bosonic degree of freedom $\ket{b} \in \left\{\ket{0}, \cdots, \ket{\La - 1}\right\}$ belong to distinct subspaces.
Using the standard vector convention for the states
\begin{align*}
  \ket{0} & = \begin{pmatrix} 1\\ 0 \end{pmatrix} &
  \ket{1} & = \begin{pmatrix} 0\\ 1 \end{pmatrix},
\end{align*}
the fermionic operators, in accordance with their action on the fermionic states in \eq{b_b_dag_on_states}, can be uniquely expressed as
\begin{align}\label{eq:b_bdag_matrix_r}
  \bhat    & = \begin{pmatrix} 0 & 1 \\ 0 & 0 \end{pmatrix} &
  \bhatdag & = \begin{pmatrix} 0 & 0 \\ 1 & 0 \end{pmatrix}.
\end{align}
The regularized bosonic position and momentum operators \qhat and \phat can act on only \La states, resulting in a matrix representation of size $(\La \times \La)$.
The matrix elements depend on the specific basis chosen.
In the next section, we explicitly represent the bosonic operators in the Fock basis and discuss their digitization for encoding onto quantum hardware.

\section{Digitization of the model}
\label{sec:fock_basis_digitization}

To implement the model on quantum hardware, the operators and degrees of freedom must be encoded using quantum gates and qubits.
For the fermion we use the well-known Jordan--Wigner transformation, which represents the fermion state as an occupation state using a single qubit. In this representation, the absence of a fermion is denoted by the state $\vert 0  \rangle $, and its presence is represented by $\vert 1 \rangle $, while the operators are
\begin{align}
  \label{eq:Jordan_Wigner}
  \bhat & = \frac{1}{2} (X +iY) &
  \bhatdag & = \frac{1}{2} (X - iY) &
  \Rightarrow \; \left[\bhatdag, \bhat\right] & = -Z,
\end{align}
where $X$, $Y$ and $Z$ represent the Pauli gates.

\begin{figure}[htpb]
\centering
\includegraphics[width=0.25\textwidth]{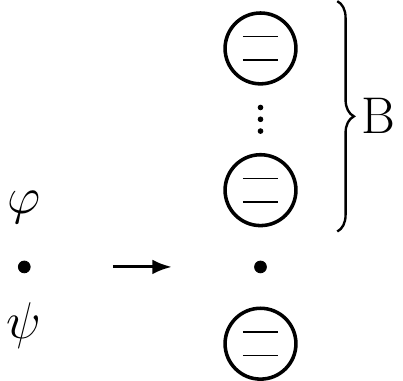}
  \caption{An illustration of the digitization of SQM. \textbf{Left:} The starting system, where the fermionic degrees of freedom is represented by $\psi$ and the bosonic one by $\phi$. \textbf{Right:} The digitization, where the symbol consisting of a circle with two lines inside represents a qubit as a two-level system. The fermionic degree of freedom requires a single qubit, while the bosonic degree of freedom is allocated $B$ qubits with $\La=2^B$ bosonic modes.}
\label{fig:qubitization_figure}
\end{figure}

The regularization of the bosonic modes, discussed in Section~\ref{sec:regularization}, is directly imposed on the matrix representation of the operators $\phat$ and $\qhat$ in the Fock basis~\cite{Sakurai_Napolitano_2020} by truncating these to \La levels:
\begin{equation}
\qhat \doteq \frac{1}{\sqrt{2 m}}
\resizebox{0.3\textwidth}{!}
{$
\left(
\begin{array}{ccccc}
0 & \sqrt{1} & 0 &\cdots &0\\
\sqrt{1} & 0 & \sqrt{2} &\cdots &0  \\
0 & \sqrt{2} & \ddots & \ddots &0  \\
0 & 0 & \ddots & 0 & \sqrt{\Lambda -1} \\
0 & 0 & \cdots & \sqrt{\Lambda -1} & 0 \\
\end{array}
\right)
$}
\hspace{1 cm}
\phat \doteq i \sqrt{\frac{m}{2}}
\resizebox{0.34\textwidth}{!}
{$
\hspace{0.04 cm}
\left(
\begin{array}{ccccc}
0 & -\sqrt{1} & 0 &\cdots &0\\
\sqrt{1} & 0 & -\sqrt{2} &\cdots &0  \\
0 & \sqrt{2} & \ddots & \ddots &0  \\
0 & 0 & \ddots & 0 & -\sqrt{\Lambda -1} \\
0 & 0 & \cdots & \sqrt{\Lambda -1} & 0 \\
\end{array}
\right)
$}
\end{equation}
For efficiency, we only consider systems where $\La = 2^B$, to make full use of the $B$ qubits allocated to the bosonic degree of freedom.
Accounting for both the bosonic and fermionic degrees of freedom, the truncated Hilbert space of SQM contains of $2\Lambda =2^{B+1}$ states and can be encoded using $B+1$ qubits.  This digitization is illustrated by Fig.~\ref{fig:qubitization_figure}.
A generic state can be written as
\begin{equation}\label{eq:state}
\vert \Psi \rangle = \vert f \rangle \vert b \rangle = \vert f \rangle  \otimes \underbrace{\vert q_{B-1} \rangle \ldots \vert q_0 \rangle}_{\substack{\text{B qubits to represent} \\ \text{the bosonic state in binary }}},
\end{equation}
and represented in quantum circuit notation as shown by \fig{fig:psi_state_circuit}.

\begin{figure}[htbp]
\centering
$\ket{\Psi}= \ket{f} \ket{b}=$
\begin{quantikz}
\lstick[4]{$\ket{b}$} &q_0  & \qw & \qw \\
&q_1 & \qw & \qw\\
&\vdots & \vdots \\
&q_{n-1} & \qw & \qw \\
\lstick[1]{$\ket{f}$} &q_{n}  & \qw & \qw \\
\end{quantikz}
\caption{An ($n + 1$)-qubit circuit for a generic state $\ket{\Psi} = \ket{f} \ket{b}$. In this representation, the first $B = n$ qubits are used to encode the $2^B$ possible bosonic states, while the last qubit provides the two possible fermionic states.}\label{fig:psi_state_circuit}
\end{figure}

\begin{figure}[htbp]
    \centering
    \begin{subfigure}{0.32\textwidth}
        \centering
        \includegraphics[width=\linewidth]{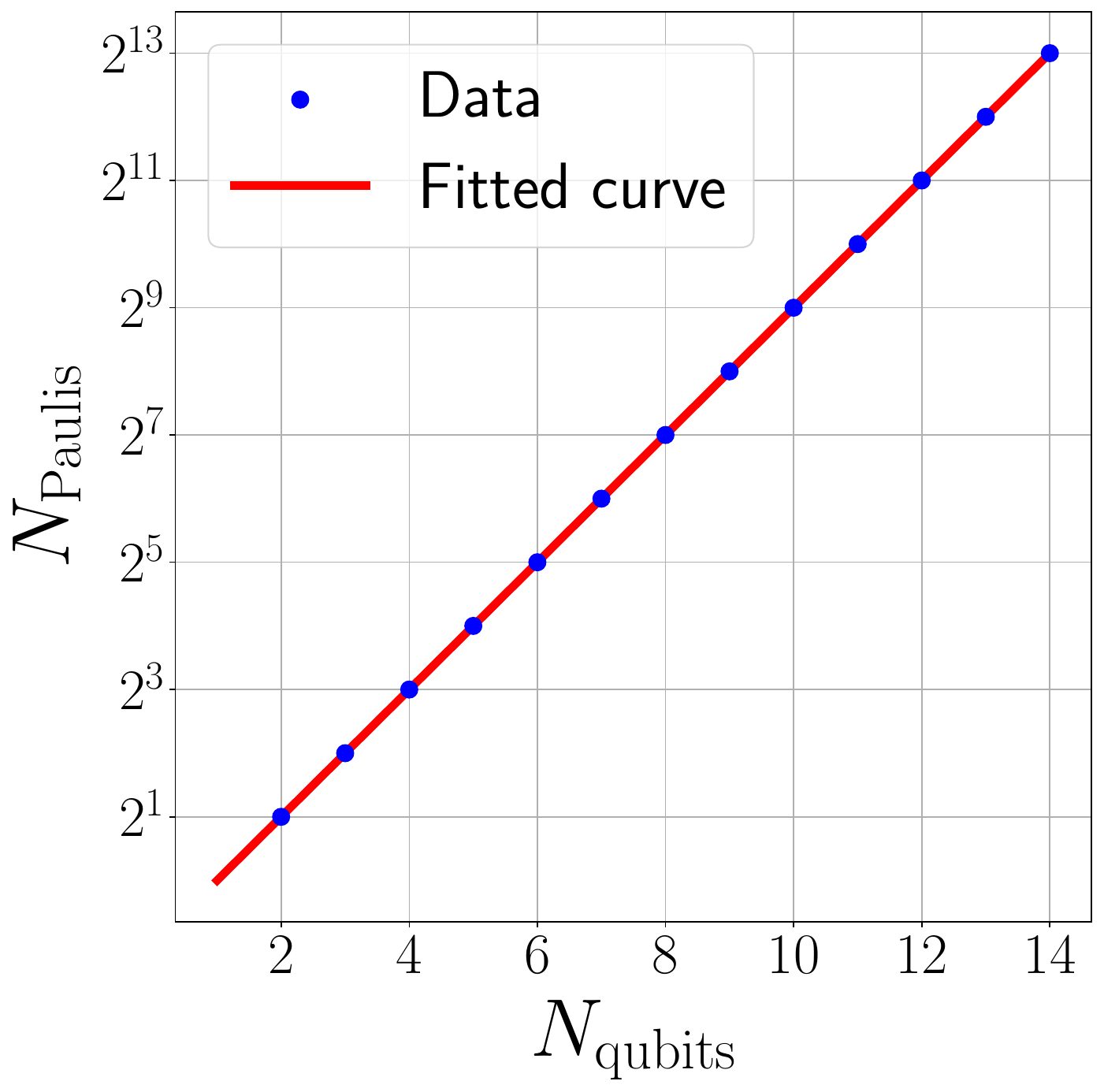}
        {Harmonic Oscillator}
    \end{subfigure}
    \hfill
    \begin{subfigure}{0.32\textwidth}
        \centering
        \includegraphics[width=\linewidth]{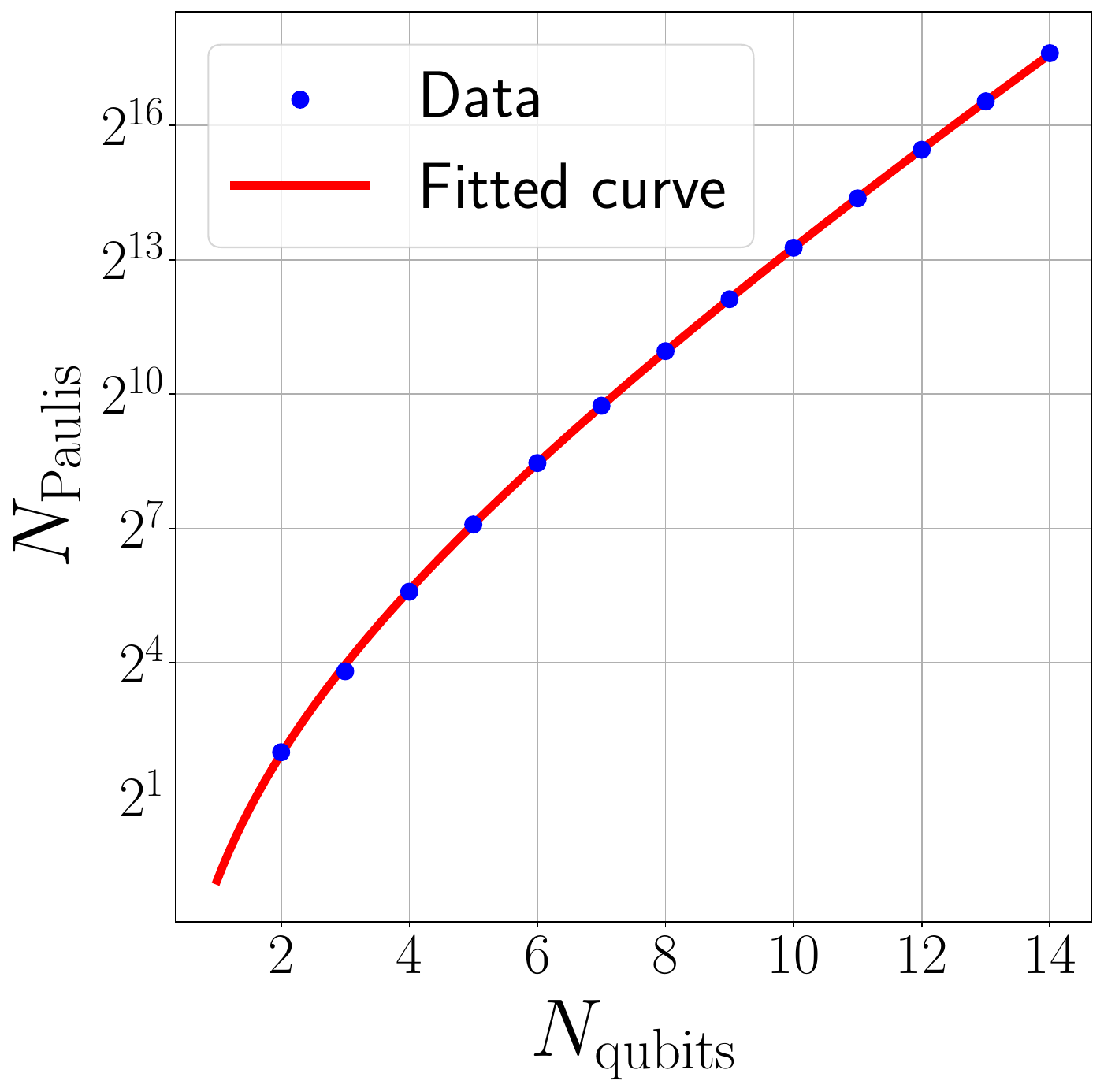}
        {Double well}
    \end{subfigure}
    \hfill
    \begin{subfigure}{0.32\textwidth}
        \centering
    \includegraphics[width=\linewidth]{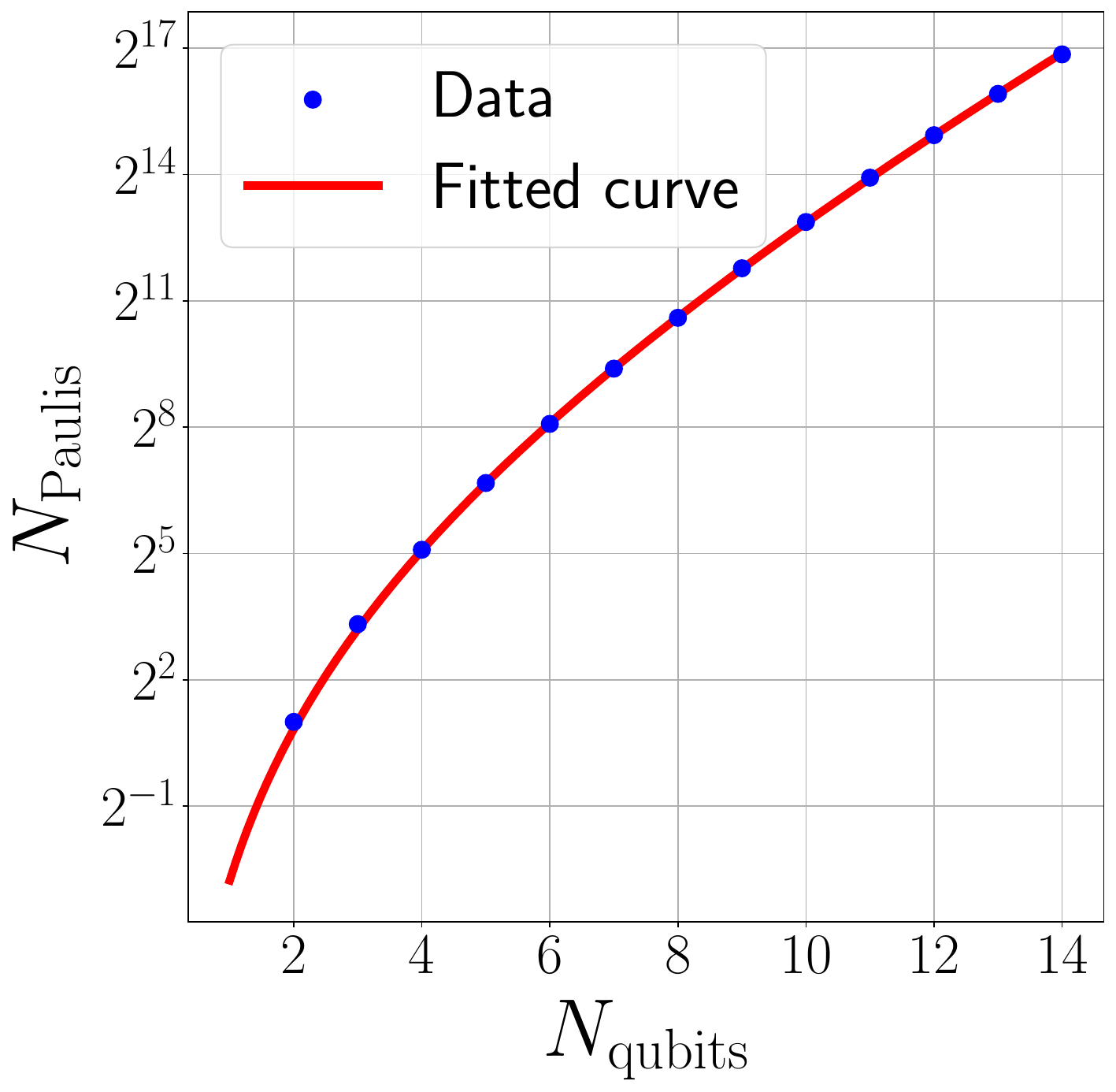}
    {Anharmonic Oscillator}
    \end{subfigure}
    \caption{The number of Pauli strings, $N_{\text{Paulis}}$, required to encode the Hamiltonian for $N_{\text{qubits}}$ qubits, corresponding to $\La = 2^{N_{\text{qubits}} - 1}$ bosonic modes.  The blue points are listed in \tab{tab:full_H_pauli_strings_energybasis}. The red curve represents the fit function $N_{\text{Paulis}} = 2^{a N_{\text{qubits}} +b}$ for the Harmonic Oscillator. For the Double Well and Anharmonic Oscillator cases, the corresponding fit is $N_{\text{Paulis}} = 2^{a N_{\text{qubits}} + c \log_2(N_{\text{qubits}}) + b}$.} \label{fig:fits_pauli_fock_basis}
\end{figure}

Once the bosonic and fermionic components of the system are digitized into qubits, the Hamiltonian is represented as a unique linear combination of Pauli strings.
We performed this transformation using the Qiskit~\cite{qiskit2024} function \texttt{SparsePauliOp}, which implements the Tensorized Pauli Decomposition algorithm proposed by Ref.~\cite{Hantzko:2023aal}. The worst-case scaling of this approach, for a system with $Q$ qubits, is $4^Q$ classical computational steps.
As shown in \tab{tab:full_H_pauli_strings_energybasis} and \fig{fig:fits_pauli_fock_basis}, the number of Pauli strings increases exponentially with the number of qubits, for each of the three superpotentials considered.
This arises from the encoded operators \phat and \qhat in the Fock basis, which were shown to scale $\propto$$Q \, 2^{Q-1}$ in Ref.~\cite{Hanada:2025yzx}.

\begin{table}[htbp]
\centering
\begin{adjustbox}{width=\textwidth}
\begin{tabular}{cccccc}
\hline
\hline
\La & $H$ size & $n$ Qubits & Harmonic Oscillator & Double Well & Anharmonic Oscillator\\ \hline
2 & $4\times 4$ & 2 & 2 & 4 & 2\\
4 & $8\times 8$ & 3 & 4 & 14 &10\\
8 & $16\times 16$ & 4 & 8 & 48 & 34\\
16 & $32\times 32$ & 5 & 16 & 136 & 102\\
32 & $64\times 64$ & 6 & 32 & 352 & 270\\
64 & $128\times 128$ & 7 & 64 & 854 & 670\\
128 & $256\times 256$ & 8 & 128 & 1990 & 1548\\
256 & $512\times 512$ & 9 & 256 & 4450 & 3496\\
512 & $1024\times 1024$ & 10 & 512 & 9874 & 7486\\
1024 & $2048\times 2048$ & 11 & 1024 & 21202 & 15534\\
2048 & $4096 \times 4096$ & 12 & 2048 &45008 & 31260 \\
4096 & $8192\times 8192$ & 13 & 4096 & 95253 & 61704\\
8192 & $16384\times 16384$ & 14 & 8192 & 200789 & 118213\\
\hline
\hline
\end{tabular}
\end{adjustbox}
\caption{A summary of the quantum resources needed to encode the full Hamiltonian using Pauli gates, for \La bosonic modes with $m = g = \mu = 1$. The number of Pauli strings required for each superpotential is listed in the corresponding column.}\label{tab:full_H_pauli_strings_energybasis}
\end{table}

In principle, this exponential scaling can be solved by adopting the position and momentum basis, where the operators \qhat and \phat become diagonal. In these representations, the encoding can be achieved using a number of Pauli gates that scales polynomially with the number of qubits. In writing the full Hamiltonian, the two representations must be connected using the quantum Fourier transform (QFT)~\cite{Coppersmith:2002skh}, an approach suggested by Refs.~\cite{Macridin:2018oli, Klco:2018zqz, Li:2022ped, Macridin:2021uwn, Hanada:2022pps, Hanada:2025yzx}.
While this approach ensures that \qhat and \phat are encoded with polynomial scaling---and consequently that the number of gates required for the Hamiltonian scales similarly---the use of the QFT introduces significant gate noise. This noise arises primarily from the CNOT gates within the QFT, and it becomes highly amplified and accumulates due to the repeated application of the QFT at each step of the variational methods used to determine both ground-state and low-lying excited-state energies. Therefore, for this study, which targets NISQ hardware, working with the Fock basis is preferable.  As we will demonstrate in the next two sections, the Hamiltonian's exponential scaling in the number of Pauli strings does not prevent us from measuring state energies with sufficiently high precision to confirm when spontaneous supersymmetry breaking occurs.

\section{Variational quantum eigensolver}
\label{sec:variational_methods}

The variational quantum eigensolver (VQE)~\cite{Peruzzo:2013bzg} is a well-known algorithm that iteratively minimizes an objective (or `cost') function.  By choosing this objective function to be the energy, the algorithm aims to find the ground state of the system. The VQE has already been demonstrated to be successful for a wide variety of systems (cf.~Ref.~\cite{Tilly:2021jem} and references therein), making it a promising means to estimate the ground-state energies of our systems.\footnote{In future work it would be interesting to also test other noise-resilient methods such as quantum imaginary time evolution~\cite{Motta:2019yya} and the quantum alternating operator ansatz~\cite{Farhi:2014ych, Hadfield:2017yqz}.} Here is a step-by-step description of the VQE algorithm:
\begin{enumerate}
    \item Prepare a parameterized quantum circuit (ansatz) $\ket{\psi(\boldsymbol{\theta})}$
    \item Measure the expectation value of $\hat{H}$ resulting in the system energy as a function of $\boldsymbol{\theta}$
    \[
    E(\boldsymbol{\theta}) = \bra{\psi(\boldsymbol{\theta})} \hat{H} \ket{\psi(\boldsymbol{\theta})}.
    \]
    \item Using a classical optimizer, vary $\boldsymbol{\theta}$ in order to minimize $E(\boldsymbol{\theta})$.
    \item Repeat steps 2 and 3 until either the energy converges or other stopping criteria are met.
\end{enumerate}
From these steps we can see there are two important things to consider when implementing this algorithm, namely the optimizer and the ansatz. We discuss each of these in more detail in the following subsections.

\subsection{Choosing an optimizer}
There are several different categories of optimizers to consider. Local, gradient-based optimizers such as Adam and L-BFGS-B \cite{lbfgs:1995,lbfgsb:1997,adam:2017} are commonly used when the cost landscape is smooth and differentiable.  These methods can be very effective when carrying out classical `statevector' simulations that exactly represent the full quantum state.  However, when using a finite number of shots to estimate expectation values (either in classical simulations or by running on a real device), the energy landscape becomes noisy and gradients can not be computed analytically. To compute gradients on a noisy device, the optimizers must employ compatible gradient evaluation methods such as the parameter shift rule and finite differences. The requirement to compute gradients in this way creates noisy gradient evaluations and becomes a significant drawback for these optimizers.

A natural next step is to turn to local, gradient-free optimizers such as COBYLA~\cite{Powell1994} or its variant COBYQA~\cite{cobyqa:rago_thesis,cobyqa:razh}.  These methods circumvent the need to calculate gradients by constructing local approximations of the objective function within a trust region --- a linear approximation for COBYLA, quadratic for COBYQA. The local nature of these methods can become an issue when the energy landscape is noisy and contains multiple local minima. It is common for these methods to get stuck in local minima and often struggle to converge to the ground state.
The prior work in Refs.~\cite{Culver:2021rxo, Culver:2023iif, Schaich:2024bmg, Mendicelli:2024ryt} employed COBYLA and observed such struggles as \La increased and the Hamiltonian became more complex, with particular challenges upon introducing shot noise instead of carrying out statevector simulations~\cite{Mendicelli:2024ryt}.

An alternative to local methods is to use stochastic, gradient-free methods such as differential evolution (DE)~\cite{DE:storn1997} or particle swarms~\cite{PSO:1995}. Differential evolution has already proven to be effective in avoiding local minima in variational quantum algorithms~\cite{Failde:2023iua, Carrascal:2023ywk}. This optimizer adopts evolutionary patterns to a population, similar to those found in nature, such as mutation, recombination, and selection. Individuals with the best traits survive longer, enabling them to produce more offspring and therefore pass on their traits to future generations. A significant drawback to these methods is the number of evaluations required at each iteration. As we increase the dimensionality of our system, we increase the total search space, resulting in the need for increasingly more circuit evaluations. This resource burden can become so large it hinders the viability of using DE with a NISQ device where run times and circuit evaluations are a limiting factor.

\begin{figure}[htbp]
    \centering
    \begin{subfigure}{\linewidth}
        \centering
        \includegraphics[width=0.9\linewidth]{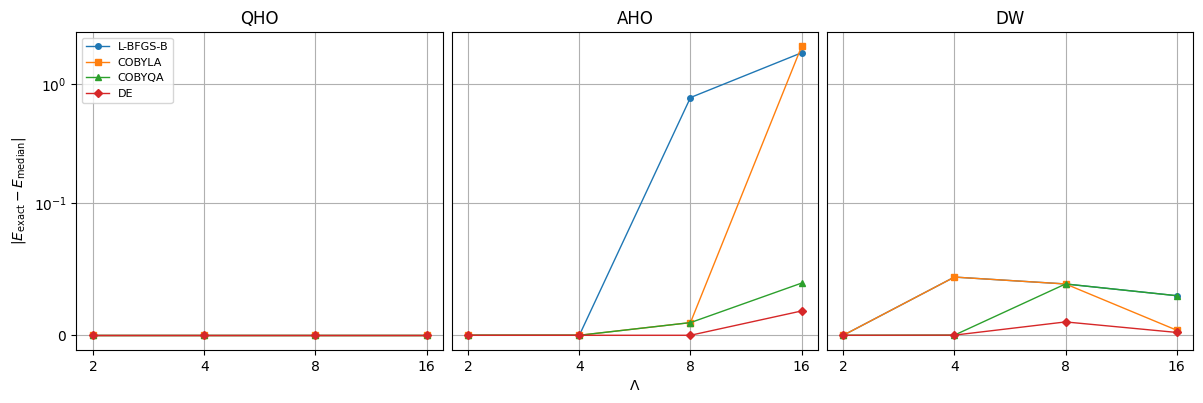}
    \end{subfigure}
    \begin{subfigure}{\linewidth}
        \centering
        \includegraphics[width=0.9\linewidth]{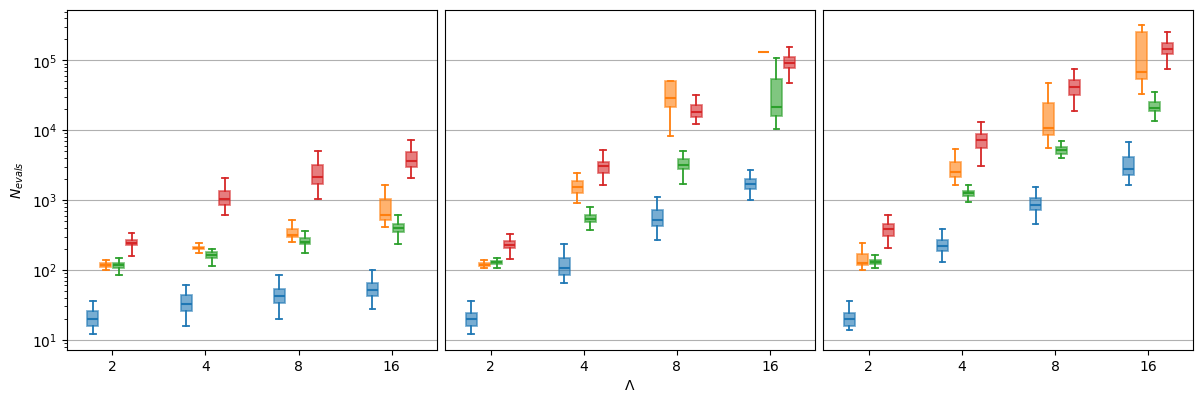}
    \end{subfigure}
    \caption{Comparison of VQE performance for various optimizers, considering 100 independent VQE runs using PennyLane's statevector simulator with a maximum of 10,000 iterations. The top plots show the absolute difference between the ground-state energy from exact diagonalization $E_{\text{exact}}$ and the median energy $E_{\text{median}}$ across the 100 independent runs for an increasing number of bosonic modes $\La$. Plotted data is from converged runs only. The bottom plots show the number of quantum circuit evaluations $N_{\text{evals}}$ aggregated over all 100 VQE runs, including runs that did not converge.}
    \label{fig:optimizer-comparison-None}
\end{figure}

\begin{figure}[htbp]
    \centering
    \begin{subfigure}{\linewidth}
        \centering
        \includegraphics[width=0.9\linewidth]{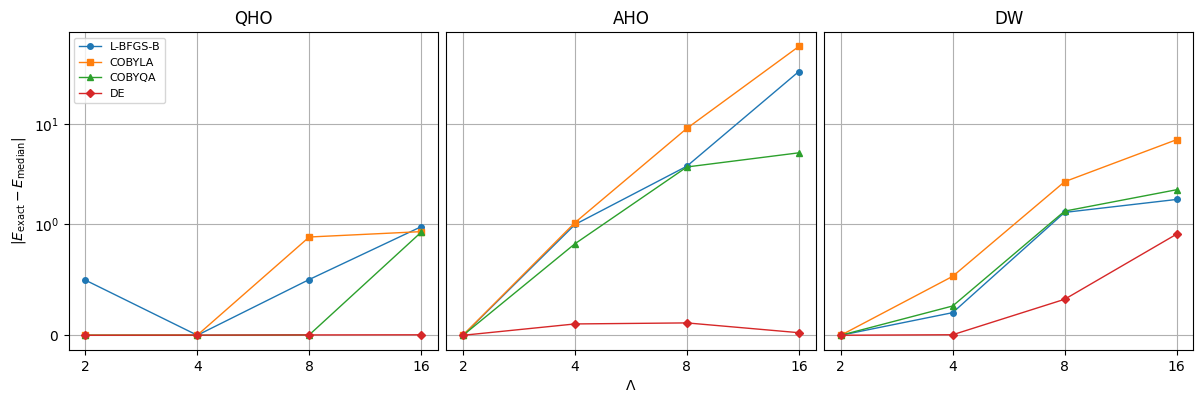}
    \end{subfigure}
    \begin{subfigure}{\linewidth}
        \centering
        \includegraphics[width=0.9\linewidth]{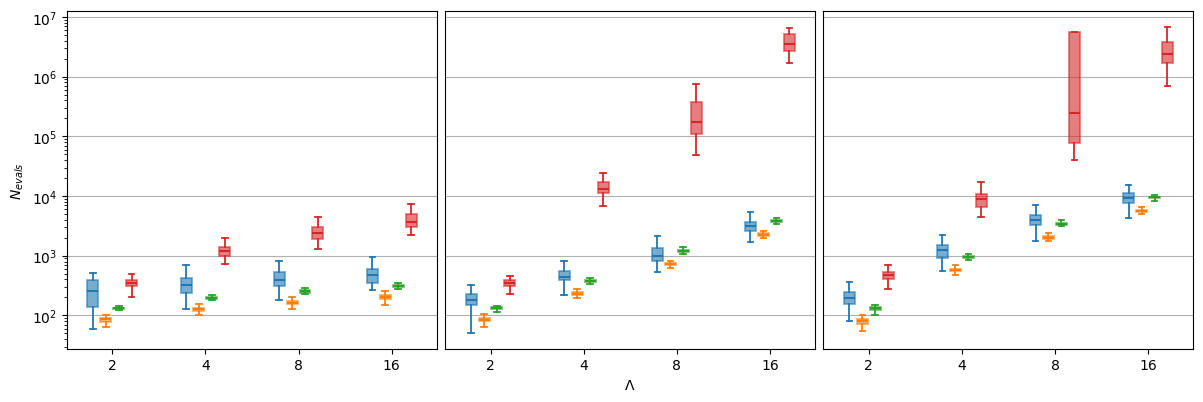}
    \end{subfigure}
    \caption{Comparison of VQE performance for various optimizers, as in \fig{fig:optimizer-comparison-None}, but now using PennyLane's default.qubit simulator with 10,000 shots.}
    \label{fig:optimizer-comparison-10k}
\end{figure}

We compare L-BFGS-B, COBYLA, COBYQA and DE in \fig{fig:optimizer-comparison-None} for statevector simulations and in \fig{fig:optimizer-comparison-10k} including shot noise (in both cases following Refs.~\cite{Culver:2021rxo, Culver:2023iif, Schaich:2024bmg, Mendicelli:2024ryt} by using a real-amplitudes ansatz discussed in the next subsection).  These figures show both how well the VQE estimates the true ground-state energy determined by exact diagonalization, as well as the total number of times the quantum circuit is evaluated, $N_{\text{evals}}$.  The latter measure accounts for the possibility of multiple evaluations per optimizer iteration, and additionally multiple evaluations as a result of decomposing the Hamiltonian into its constitute Pauli terms before measurement. $N_{\text{evals}}$ is not inclusive of the number of shots which would simply scale each point in \fig{fig:optimizer-comparison-10k} by a constant factor. We choose to use the median energy across 100 VQE runs as our metric for gauging how well the ground-state energy is estimated.  Although the minimum energies across these runs may come closer to $E_{\text{exact}}$, the median is more robust to outliers, especially when considering noisy systems.

We can clearly see the trade-off we have to carefully consider when using DE. In terms of accuracy, this method clearly outperforms the other optimizers. However, for DE to provide a good approximation it requires a larger number of evaluations per iteration. The minimum population size we can use is  $5\times\text{N}_{\text{params}}$. Since here we consider the real-amplitudes ansatz, the number of variational parameters scales with the number of qubits and therefore the population size will also scale alongside this. This results in a significant increase in the number of circuit evaluations $N_{\text{evals}}$ as we increase $\La$.  For statevector simulations with $\La \leq 16$, the number of evaluations required for DE can remain comparable to that needed by local methods.  With shot noise, however, DE is orders of magnitude more expensive, with $\sim$$10^6$--$10^7$ circuit evaluations needed for $\La=16$. This number of evaluations can be reduced through methods such as grouping the Pauli strings into commutable terms before measurement, however it is still not feasible to run this many evaluations on a NISQ device.

Another notable feature of these figures are cases in which local optimizers require more evaluations for statevector compared to noisy simulation. An example of this is the COBYQA optimizer for the AHO superpotential with $\La=8$. This is due to the fact that when we introduce noise we make the energy landscape more complex and it is more likely for the optimizer to get stuck in local minima. Note that the median energy ends up farther from the ground-state value when the number of evaluations decreases in the presence of shot noise. When we use statevector simulation the energy landscape is less complex and the optimizer is able to converge closer to the ground state, at the expense of a larger number of iterations.

Since we are looking forward to running these circuits on a real NISQ device, we use the COBYQA optimizer for the remainder of this paper.  Although DE performs the best in terms of accuracy, the number of evaluations it requires prevents us from realistically using it on a NISQ device. While L-BFGS-B is a good choice for statevector simulations, its dependence on computing gradients is problematic when considering using a NISQ device.  In addition to reducing the overall number of evaluations needed during the optimization process, we can also work on simplifying and speeding up these circuit evaluations.  In particular, for all optimizers, we can see the significant costs of employing an ansatz with a number of variational parameters that scales with the number of qubits.  This motivates work to design a more efficient ansatz, to which we now turn.

\subsection{Choosing an ansatz}
In heuristic methods, choosing the ansatz --- the initial educated guess for the state candidate --- is one of the critical steps in ensuring the successful convergence of the quantum variational method to the correct solution.  While we are currently considering systems for which the solution can be found by classical exact diagonalization, our ultimate aim is to go beyond this regime and address problems that are classically intractable.  Without detailed information about the ground state, hardware-efficient ans\"atze are often employed. There are many different types of hardware-efficient ans\"atze, as outlined in Ref.~\cite{HardwareEfficent:2019}. One of the most commonly used is the real-amplitudes ansatz mentioned in the previous subsection, which itself comes in many variations that alter the way entanglement is implemented. As the name suggests, these ans\"atze are designed to be device agnostic, with the gates being largely applicable and well calibrated to most devices. This positions them well for general-purpose use as they offer great flexibility and many can be easily manipulated to respect symmetries and boundary conditions of a problem.

A drawback to these ans\"atze is the fact that the number of gates and variational parameters they involve increases rapidly with the number of qubits. Since we need to increase the number of qubits in order to extrapolate $\La \to \infty$ and remove the explicit supersymmetry breaking discussed above, using an ansatz that scales with the number of qubits becomes problematic~\cite{Schaich:2024bmg, Mendicelli:2024ryt}, especially when using NISQ devices. Additionally, since these ans\"atze typically treat all qubits the same, it is likely that many of the gates and parameters they involve are not actually required. In the NISQ era, the number of quantum gates, and in particular two-qubit entangling gates, needs to be carefully managed. Gate errors will affect the accuracy of the ansatz initialization and therefore affect the overall optimization process. Ideally we want to keep gate counts to a minimum, which motivates us to look beyond general-purpose ans\"atze.

This issue has sparked recent research into systematic methods of creating problem-tailored ans\"atze using more adaptive approaches, including Refs.~\cite{Grimsley:2018wnd, Grimsley:2022azc, Farrell:2023fgd, Farrell:2024fit, farrell2025dqss, Gustafson:2024bww, Shirali:2025cox}.  We now present our approach to construct more efficient ans\"atze that can be extrapolated to general values of $\La$.

\subsubsection{Adaptive VQE ansatz construction}\label{section:Adaptive-VQE}
Our aim is to produce problem-tailored ans\"atze that have fewer gates compared to (e.g.) the real-amplitudes ans\"atze used by Refs.~\cite{Schaich:2024bmg, Mendicelli:2024ryt}, while still being expressive enough to produce the ground state. This is a common challenge that variational algorithms face and there has already been extensive research into this area. The Adaptive Derivative-Assembled Problem Tailored (ADAPT) VQE, and similar methods based on it, have been successfully applied to similar problems~\cite{Grimsley:2018wnd, Grimsley:2022azc, Farrell:2023fgd, Farrell:2024fit, farrell2025dqss, Gustafson:2024bww, Shirali:2025cox}. These algorithms iteratively construct an ansatz by selecting an operator to add at each step, taken from from a pre-defined pool. An operator is selected based how it affects the energy, for example by computing the largest-magnitude gradient of the expectation value, the infidelity, or the commutator of the Hamiltonian with the operator. This ensures operators that have the largest effect on the energy are included first and prevents unnecessary gates from being added.

The ADAPT-VQE algorithm incorporates operators $\hat O_i$ into the ansatz wavefunction via parameterized evolution, $\ket{\psi_{\text{ansatz}}} \to e^{i\theta_i \hat O_i} \ket{\psi_{\text{ansatz}}}$, where $\theta_i$ is the variational parameter to be optimized.
The Scalable-Circuits (SC-ADAPT-VQE) variant~\cite{Farrell:2023fgd, Farrell:2024fit} builds on this by exploiting locality and translation invariance in gapped systems to scale ansatz construction from small problem sizes (within reach of classical simulation) to larger ones that would be prohibitively expensive to tackle directly.
A further variant, the Surrogate-Constructed (SC)$^2$-ADAPT-VQE~\cite{Gustafson:2024bww}, exploits a classically calculable `surrogate' approximation of the ground state to truncate the operator pool and thereby stabilize the extrapolation to large systems.

Taking inspiration from these methods, we have developed our own approach, which we will simply refer to as an Adaptive VQE (AVQE) algorithm. This method contains the same core ideas as the ADAPT-VQE algorithm, however instead of using parameterized evolution we define our operator pool to consist of rotation gates and controlled rotation gates, with the rotation angles providing our variational parameters.
Operator selection is determined by calculating the largest-magnitude gradient for the expectation value of the Hamiltonian.  Specifically, the AVQE algorithm is performed as follows:
\begin{enumerate}
    \item Define an operator pool \{$\hat{O}$\} as discussed further below.
    \item Initialize the ansatz state $\ket{\psi_{\text{ansatz}}}$ in a chosen basis state.
    \item Measure the gradient of the expectation value of the Hamiltonian for each operator in the pool:
        \begin{equation}\label{eq:ADAPT-VQE}
            \nabla\bra{\psi_{\text{ansatz}}}\hat{O}_{i}(\theta)^\dag\hat{H}\hat{O}_{i}(\theta)\ket{\psi_{\text{ansatz}}}
        \end{equation}
    \item Add to the ansatz the operator that returns the largest gradient.
    \item Perform a VQE using the updated ansatz. If an operator $\hat O_i$ has been optimized in a previous VQE step then use the previous optimal parameter as the initial rotation angle $\theta_i$. For the new operator $\hat O_n$ initialize $\theta_n=0$.
    \item If the change in energy is below a pre-determined threshold then terminate the algorithm. Else return to step 3 with the updated ansatz.
\end{enumerate}

Although computing gradients on a noisy device can be problematic, our plan is to carry out this adaptive ansatz construction for small systems where statevector simulation is feasible. By identifying patterns in the ans\"atze produced for these smaller, tractable systems, we can then propose efficient problem-tailored ans\"atze suitable for extrapolation to larger \La out of reach of classical statevector simulation.

The first steps of the AVQE algorithm above involve determining a suitable operator pool and basis state.
Choosing a suitable operator pool is vital and there are several things to consider when constructing a pool. We aim to minimize the number of operators in the pool so that we reduce the number of times we are required to calculate \eq{eq:ADAPT-VQE}. That said, we must also ensure the pool is diverse and expressive enough to be capable of producing the required ground state. For the SQM systems we consider here, we find the following set of rotation operators is sufficient and effective:
\begin{equation}
    \{{\hat{O}}\} = \{RY, RZ, CRY\}.
\end{equation}
These gates form a complete set, which in combination have the capability to perform any arbitrary rotation. The inclusion of the controlled CRY gate enables entanglement to arise within the circuit.

Choosing a suitable initial state can also be another crucial step. The choice of initial state is often inspired by some symmetry of the system or prior knowledge that the ground state lies in a particular subspace. In the case of SQM there is a clear block structure to the Hamiltonian which is outlined in more detail in Appendix~\ref{app:block-structure}. The ground state is associated with one of two blocks that depend on the state of the fermion, enabling the system to be reduced to that block.  From direct inspection of energy eigenvectors for small $\La$, we can see the ground state for the HO and AHO superpotentials has the fermion qubit in the $\ket{1}$ state. Similarly, for the DW superpotential with $\La \geq 8$ the ground state has the fermion qubit in the $\ket{0}$ state. We incorporate these observations into our initial states for arbitrary $\La$. In principle, we could reduce our Hamiltonian to just the single block containing the ground state, and completely remove the fermion qubit from the setup. However, we expect this is possible only for SQM, and choose to retain the whole Hamiltonian in preparation for moving to more complex systems such as the $(1+1)$-dimensional Wess--Zumino model. Still, this possibility does highlight the significance of choosing a good initial state.

In summary, our initial states are $\ket{10\dots0}$ for the HO and AHO superpotentials, and $\ket{00\dots0}$ for the DW superpotential with $\La \neq 4$. These basis states are represented in Qiskit's little-endian notation, i.e., $\ket{q_n, \dots, q_1, q_0}$ where $q_n$ is the qubit for the fermion (cf.~\fig{fig:psi_state_circuit}). Utilizing these basis states places the optimizer in a preferable search space, making it significantly easier for the algorithm to find the correct ground state. In fact, an incorrect choice of initial state could often mean the difference between the algorithm converging to the correct ground state or not.

\subsubsection{Generalizing the AVQE ansatz to arbitrary $\La$}
\label{ansatz_extrapolation}

\begin{figure}[htbp]
    \centering
    \includegraphics[width=0.9\textwidth]{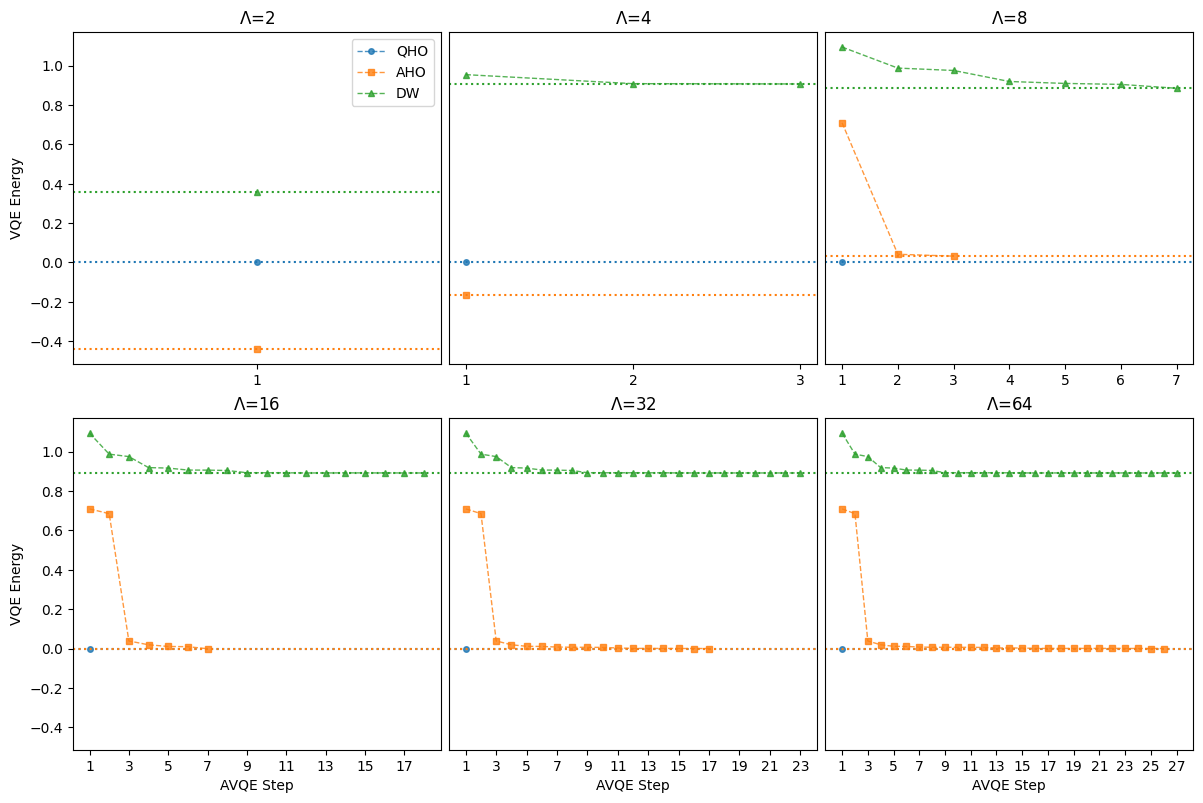}
    \caption{AVQE energies at each step of the algorithm for each superpotential and increasing values of $\La$, using PennyLane's default.qubit statevector simulation and the COBYQA optimizer. Dashed lines indicate the minimum eigenvalue from exact diagonalization.}
    \label{fig:avqe-steps}
\end{figure}

\tab{tab:AVQE-Summary} summarizes AVQE ansatz construction results for each superpotential with \La ranging from 2 to 64.
In each case we run the algorithm 100 times with different random initialization of the classical COBYQA optimizer, and take the best ansatz to be the one that returns a VQE energy closest to the exact value.
It is possible that better ans\"atze may exist but be missed by this procedure.
This becomes increasingly likely as we consider larger \La where the optimization becomes more challenging.
As we can see in \fig{fig:avqe-steps}, at larger \La the VQE energy drops quickly during the first few steps, but then improves much more slowly. This may suggest the optimizer is struggling to converge, and the ground-state energy may be smaller than what the algorithm finds. We can also see in \tab{tab:AVQE-Summary} that as we get to larger \La the precision of the VQE energy begins to decline. Again, this is likely a result of the classical optimizer struggling to resolve the increasing number of variational parameters alongside an increasingly complex Hamiltonian.

\begin{table}[htbp]
\centering
\resizebox{\textwidth}{!}{%
\begin{tabular}{lrlclll}
\toprule
$W(\qhat)$ & \La & Basis State & $N_\text{gates}$ & Ansatz & $E_{\text{VQE}}$ & $E_{\text{Exact}}$ \\
\midrule
HO & 2 & $\ket{10}$ & 1 & RY[$q_1$] & 0.000000 & 0.000000 \\
HO & 4 & $\ket{100}$ & 1 & RY[$q_2$] & 0.000000 & 0.000000 \\
HO & 8 & $\ket{1000}$ & 1 & RY[$q_3$] & 0.000000 & 0.000000 \\
HO & 16 & $\ket{10000}$ & 1 & RY[$q_4$] & 0.000000 & 0.000000 \\
HO & 32 & $\ket{100000}$ & 1 & RY[$q_5$] & 0.000000 & 0.000000 \\
HO & 64 & $\ket{1000000}$ & 1 & RY[$q_6$] & 0.000000 & 0.000000 \\
$\vdots$&&&&&&\\
HO& \La & $\ket{10\dots0}$ & 1 & RY[$q_n$] &&\\
\midrule
AHO & 2 & $\ket{10}$ & 1 & RY[$q_1$] & -0.437500 & -0.437500 \\
AHO & 4 & $\ket{100}$ & 1 & RY[$q_1$] & -0.164785 & -0.164785 \\
AHO & 8 & $\ket{1000}$ & 3 & RY[$q_2$], RY[$q_1$], CRY[$q_1$, $q_2$] & 0.032010 & 0.032010 \\
AHO & 16 & $\ket{10000}$ & 7 & RY[$q_2$], RY[$q_3$], RY[$q_1$], CRY[$q_1$, $q_2$], CRY[$q_1$, $q_3$], CRY[$q_2$, $q_3$], CRY[$q_3$, $q_2$] & -0.001167 & -0.001167 \\
AHO & 32 & $\ket{100000}$ & 24 & RY[$q_2$], RY[$q_3$], RY[$q_1$], CRY[$q_1$, $q_2$], CRY[$q_1$, $q_3$], RY[$q_4$], CRY[$q_2$, $q_3$] $\dots$ , RY[$q_3$] & 0.000006 & 0.000006 \\
AHO & 64 & $\ket{1000000}$ & 28 & RY[$q_2$], RY[$q_3$], RY[$q_1$], CRY[$q_1$, $q_2$], CRY[$q_1$, $q_3$], RY[$q_4$], CRY[$q_2$, $q_3$] $\dots$  , RY[$q_4$] & 0.000019 & -0.000000 \\
$\vdots$&&&&&&\\
AHO & \La & $\ket{10\dots0}$ & 4 & RY[$q_2$], RY[$q_3$], RY[$q_1$], CRY[$q_1$, $q_2$] &&\\
\midrule
DW & 2 & $\ket{00}$ & 1 & RY[$q_0$] & 0.357233 & 0.357233 \\
DW & 4 & $\ket{100}$ & 3 & RY[$q_1$], RY[$q_0$], CRY[$q_1$, $q_0$] & 0.906560 & 0.906560 \\
DW & 8 & $\ket{0000}$ & 7 & RY[$q_0$], CRY[$q_0$, $q_1$], RY[$q_2$], RY[$q_1$], CRY[$q_0$, $q_2$], CRY[$q_1$, $q_2$], RY[$q_1$] & 0.884580 & 0.884580 \\
DW & 16 & $\ket{00000}$ & 18 & RY[$q_0$], CRY[$q_0$, $q_1$], RY[$q_2$], RY[$q_1$], RY[$q_3$], CRY[$q_0$, $q_2$], CRY[$q_0$, $q_3$] $\dots$ , RY[$q_2$] & 0.891599 & 0.891599 \\
DW & 32 & $\ket{000000}$ & 23 & RY[$q_0$], CRY[$q_0$, $q_1$], RY[$q_2$], RY[$q_1$], RY[$q_3$], CRY[$q_0$, $q_2$], CRY[$q_0$, $q_3$] $\dots$ , RY[$q_3$] & 0.891634 & 0.891632 \\
DW & 64 & $\ket{0000000}$ & 27 & RY[$q_0$], CRY[$q_0$, $q_1$], RY[$q_2$], RY[$q_1$], RY[$q_3$], CRY[$q_0$, $q_2$], CRY[$q_0$, $q_3$] $\dots$ , RY[$q_5$] & 0.891635 & 0.891632 \\
$\vdots$&&&&&&\\
DW & \La & $\ket{00\dots0}$ & 4 & RY[$q_0$], CRY[$q_0$, $q_1$], RY[$q_2$], RY[$q_1$] &&\\
\bottomrule
\end{tabular}
}
 \caption{Summary of 100 AVQE results for each superpotential and increasing number of bosonic modes $\La$, using statevector simulation and the COBYQA optimizer. In the basis state the left-most bit corresponds to the fermion and the following bits represent the boson. The rotation gates in the `Ansatz' column act on the qubit(s) specified by the number(s) in the brackets. In the case of CRY the control and target qubits are [control,target]. The final two columns compare the best ansatz VQE energy from the 100 runs and the true ground-state energy from exact diagonalization. The final row for each superpotential shows the generalized ansatz extrapolated to arbitrary $\La$. This involves truncating the circuits to include only the first four operators, as discussed in the text.}
 \label{tab:AVQE-Summary}
\end{table}

Looking at the `Ansatz' column in \tab{tab:AVQE-Summary} we can see how the ansatz is systematically constructed. For the trivial HO superpotential, only a single RY gate on the fermion ($q_n$) is required. This is no surprise since this model is non-interacting and we are representing our boson in the Fock basis. This is the case for all cutoffs and can be easily extrapolated to arbitrary $\La$.
There is also a clear pattern in the first few gates added to the ans\"atze for the AHO and DW superpotentials.  Although the total number of gates in the ansatz, $N_{\text{gates}}$, increases with $\La$, we can see the same initial gates are added for all $\La \ge 16$. Since by design the algorithm first adds the gates that have the most impact on the energy, we can infer that these initial gates are the most important and subsequent gates have diminishing effects. To confirm this we have plotted the AVQE energies at each step in \fig{fig:avqe-steps}.

Utilizing more gates and parameters increases the expressivity of the ansatz, potentially allowing for a more accurate approximation of the ground state.  However this also makes the energy landscape more complex. This added complexity comes at a price. On NISQ devices, an increase in circuit depth means an increase in errors and as a result there becomes a point where adding more gates to the circuit has a negative impact on the VQE energy. This can be seen in \tab{tab:AVQE-Steps}. When using statevector simulation, in the AVQE steps we can see a clear improvement in the VQE energy at each step, with eventual convergence to the exact ground-state energy. However, when we run the VQE using the resulting ans\"atze, with 10,000 shots rather than statevector simulation, we do not observe the same improvement. In fact, we observe a decline in accuracy as the number of gates increases. Accounting for device noise, decoherence and gate errors, in addition to shot noise, is expected to make this problem even worse.

Looking at $\Delta E$ for the 10,000-shots VQE in \tab{tab:AVQE-Steps} we can see a steady improvement for the first four gates. After this, performance tends to decline as a result of the added complexity introduced by increasing the number of gates.  After these four gates, the statevector simulation in the AVQE ansatz construction produces $\Delta E \simeq 10^{-2}$. Although not perfect, this difference is negligible compared to the effects of shot noise, making it acceptable to truncate the ansatz circuit after four gates.  This truncated ansatz is what we extrapolate to larger $\La$. The generalized ans\"atze are shown in the final row for each superpotential in \tab{tab:AVQE-Summary}. Truncating the ans\"atze of course comes with the concern that we are intentionally reducing, or even removing, the ability of the VQE to find the ground state. Although this is the case, these first four gates have the greatest impact on the energy, and offer the best chance of distinguishing zero vs.\ non-zero ground-state energy.  The expressivity of the ansatz needs to be balanced against the computational complexity in order to obtain the best performance.

\begin{figure}[htbp]
    \centering
    \begin{subfigure}{0.25\textwidth}
        \centering
        \includegraphics[width=\linewidth]{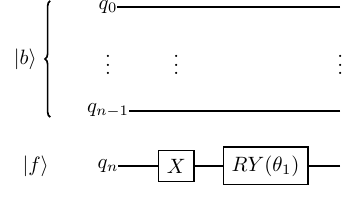}
        {Harmonic Oscillator}
    \end{subfigure}
    \hfill
    \begin{subfigure}{0.35\textwidth}
        \centering
        \includegraphics[width=\linewidth]{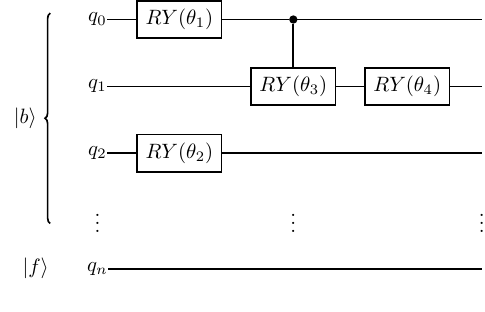}
        {Double well}
    \end{subfigure}
    \hfill
    \begin{subfigure}{0.25\textwidth}
        \centering
    \includegraphics[width=\linewidth]{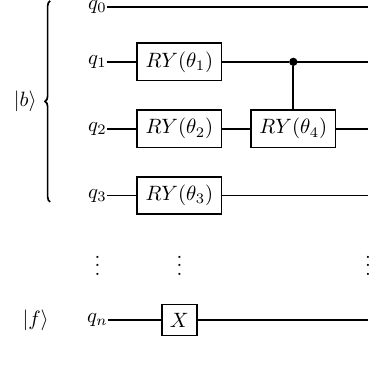}
    {Anharmonic Oscillator}
    \end{subfigure}
    \caption{Circuit diagrams for the general-$\La$ ans\"atze tailored for each of the three superpotentials, obtained by applying the AVQE algorithm to systems with an increasing number of bosonic modes $\La$, followed by the truncation discussed in the text.}
    \label{fig:ansatz_circuit_c_limit}
\end{figure}

We have now reached our goal of proposing a suitable ansatz for large \La by extrapolating those obtained from the AVQE algorithm for smaller-$\La$ systems.  \Cref{fig:ansatz_circuit_c_limit} shows the circuit diagrams for our new general-$\La$ ans\"atze tailored for each superpotential.  The number of gates and variational parameters in these ans\"atze no longer increases with the number of qubits, in contrast to the real-amplitudes ans\"atze used by Refs.~\cite{Schaich:2024bmg, Mendicelli:2024ryt}.  This gives us access to larger \La than was previously feasible and is especially important for NISQ devices, where coherence times and circuit depth play a large role in performance.

\begin{table}[htbp]
\centering
\resizebox{0.9\textwidth}{!}{%
\begin{tabular}{lccl|cc|cc|}
\cline{5-6} \cline{7-8}
\multicolumn{4}{l|}{} & \multicolumn{2}{c|}{AVQE} & \multicolumn{2}{c|}{VQE-10k Shots} \\
\hline
$W(\qhat)$ & \La &  Step & Ansatz   &   $E_{\text{VQE}}$ & $\Delta E$ &  $E_{\text{VQE}}$ & $\Delta E$\\
\hline
AHO & 2 & 1 & RY[$q_1$] & -0.437500 & 0.000000 & -0.437500 & 0.000000 \\
\hline
AHO & 4 & 1 & RY[$q_1$] & -0.164785 & 0.000000 & -0.227023 & 0.062238 \\
\hline
AHO & 8 & 1 & RY[$q_2$] & 0.709178 & 0.677168 & -0.030801 & 0.062811 \\
AHO & 8 & 2 & RY[$q_2$], RY[$q_1$] & 0.041604 & 0.009594 & -0.048969 & 0.080979 \\
AHO & 8 & 3 & RY[$q_2$], RY[$q_1$], CRY[$q_1$, $q_2$] & 0.032010 & 0.000000 & -0.009415 & 0.041425 \\
\hline
AHO & 16 & 1 & RY[$q_2$] & 0.709178 & 0.710345 & 1.871556 & 1.872723 \\
AHO & 16 & 2 & RY[$q_2$], RY[$q_3$] & 0.684881 & 0.686048 & 1.129883 & 1.131050 \\
AHO & 16 & 3 & RY[$q_2$], RY[$q_3$], RY[$q_1$] & 0.038341 & 0.039508 & 1.051022 & 1.052189 \\
AHO & 16 & 4 & RY[$q_2$], RY[$q_3$], RY[$q_1$], CRY[$q_1$, $q_2$] & 0.018906 & 0.020073 & 0.656795 & 0.657962 \\
AHO & 16 & 5 & RY[$q_2$], RY[$q_3$], RY[$q_1$], CRY[$q_1$, $q_2$], CRY[$q_1$, $q_3$] & 0.010364 & 0.011531 & 1.468272 & 1.469439 \\
AHO & 16 & 6 & RY[$q_2$], RY[$q_3$], RY[$q_1$], CRY[$q_1$, $q_2$], CRY[$q_1$, $q_3$], CRY[$q_2$, $q_3$] & 0.009265 & 0.010432 & 8.471542 & 8.472709 \\
AHO & 16 & 7 & RY[$q_2$], RY[$q_3$], RY[$q_1$], CRY[$q_1$, $q_2$], CRY[$q_1$, $q_3$], CRY[$q_2$, $q_3$], CRY[$q_3$, $q_2$] & -0.001167 & 0.000000 & 6.378891 & 6.380058 \\
\hline
\\
\hline
DW & 2 & 1 & RY[$q_0$] & 0.357233 & 0.000000 & 0.357233 & 0.000000 \\
\hline
DW & 4 & 1 & RY[$q_0$] & 0.953903 & 0.047343 & 0.927232 & 0.020672 \\
DW & 4 & 2 & RY[$q_0$], RY[$q_1$] & 0.908098 & 0.001538 & 1.024910 & 0.118350 \\
DW & 4 & 3 & RY[$q_0$], RY[$q_1$], CRY[$q_1$, $q_0$] & 0.906560 & 0.000000 & 1.000301 & 0.093741 \\
\hline
DW & 8 & 1 & RY[$q_0$] & 1.093911 & 0.209331 & 0.826379 & 0.058202 \\
DW & 8 & 2 & RY[$q_0$], CRY[$q_0$, $q_1$] & 0.987155 & 0.102575 & 1.758564 & 0.873984 \\
DW & 8 & 3 & RY[$q_0$], CRY[$q_0$, $q_1$], RY[$q_2$] & 0.975092 & 0.090511 & 1.669800 & 0.785219 \\
DW & 8 & 4 & RY[$q_0$], CRY[$q_0$, $q_1$], RY[$q_2$], RY[$q_1$] & 0.919089 & 0.034508 & 1.587420 & 0.702840 \\
DW & 8 & 5 & RY[$q_0$], CRY[$q_0$, $q_1$], RY[$q_2$], RY[$q_1$], CRY[$q_0$, $q_2$] & 0.909293 & 0.024712 & 2.889190 & 2.004610 \\
DW & 8 & 6 & RY[$q_0$], CRY[$q_0$, $q_1$], RY[$q_2$], RY[$q_1$], CRY[$q_0$, $q_2$], CRY[$q_1$, $q_2$] & 0.904376 & 0.019795 & 2.313434 & 1.428854 \\
DW & 8 & 7 & RY[$q_0$], CRY[$q_0$, $q_1$], RY[$q_2$], RY[$q_1$], CRY[$q_0$, $q_2$], CRY[$q_1$, $q_2$], RY[$q_1$] & 0.884580 & 0.000000 & 2.469986 & 1.585405 \\
 \hline
DW & 16 & 1 & RY[$q_0$] & 1.093911 & 0.202312 & 0.507588 & 0.384012 \\
DW & 16 & 2 & RY[$q_0$], CRY[$q_0$, $q_1$] & 0.987155 & 0.095556 & 0.456968 & 0.434632 \\
DW & 16 & 3 & RY[$q_0$], CRY[$q_0$, $q_1$], RY[$q_2$] & 0.974805 & 0.083206 & 0.459035 & 0.432564 \\
DW & 16 & 4 & RY[$q_0$], CRY[$q_0$, $q_1$], RY[$q_2$], RY[$q_1$] & 0.918832 & 0.027233 & 0.761203 & 0.130396 \\
DW & 16 & 5 & RY[$q_0$], CRY[$q_0$, $q_1$], RY[$q_2$], RY[$q_1$], RY[$q_3$] & 0.916445 & 0.024846 & 0.876312 & 0.015287 \\
DW & 16 & 6 & RY[$q_0$], CRY[$q_0$, $q_1$], RY[$q_2$], RY[$q_1$], RY[$q_3$], CRY[$q_0$, $q_2$] & 0.906359 & 0.014760 & 1.366512 & 0.474913 \\
DW & 16 & 7 & RY[$q_0$], CRY[$q_0$, $q_1$], RY[$q_2$], RY[$q_1$], RY[$q_3$], CRY[$q_0$, $q_2$], CRY[$q_0$, $q_3$] & 0.905997 & 0.014398 & 2.540056 & 1.648457 \\
\vdots &&&&&&&\\
DW & 16 & 18 & RY[$q_0$], CRY[$q_0$, $q_1$], RY[$q_2$], RY[$q_1$], RY[$q_3$] \dots, CRY[$q_3$, $q_1$], RY[$q_2$] & 0.891599 & 0.000000 &  5.144443 & 4.252844 \\
\hline
\end{tabular}
}
 \caption{VQE results for each step of the AVQE algorithm for the DW and AHO superpotentials. Each step adds a single gate as shown in the Ansatz column, with the same notation as in \tab{tab:AVQE-Summary}.  $E_{\text{VQE}}$ is the resulting VQE energy at each step and $\Delta E$ is the difference to the ground-state energy from exact diagonalization. The `AVQE' results are the minimum $\Delta E$ across 100 independent AVQE runs using statevector simulation. The `VQE-10K Shots' results are the median from 100 independent VQE runs using 10,000 shots and the given ansatz at each step.}
 \label{tab:AVQE-Steps}
\end{table}

\subsection{Results}
We conclude this section by collecting our VQE results for SQM with each of the HO, AHO and DW superpotentials, comparing three different ans\"atze, with and without shot noise.
All of these calculations use the COBYQA optimizer.
The ans\"atze we compare in each case are (i) a real-amplitudes ansatz like those used by Refs.~\cite{Culver:2021rxo, Culver:2023iif, Schaich:2024bmg, Mendicelli:2024ryt}; (ii) the ``Full'' ansatz returned by the AVQE algorithm; and (iii) the AVQE ansatz ``Truncated'' to include at most the first four gates.
\tab{tab:AVQE-Summary} summarizes both the Full and Truncated ans\"atze, the latter of which are also shown in \fig{fig:ansatz_circuit_c_limit}.

\begin{figure}[htbp]
    \centering
    \begin{subfigure}{\linewidth}
        \centering
        \includegraphics[width=0.9\linewidth]{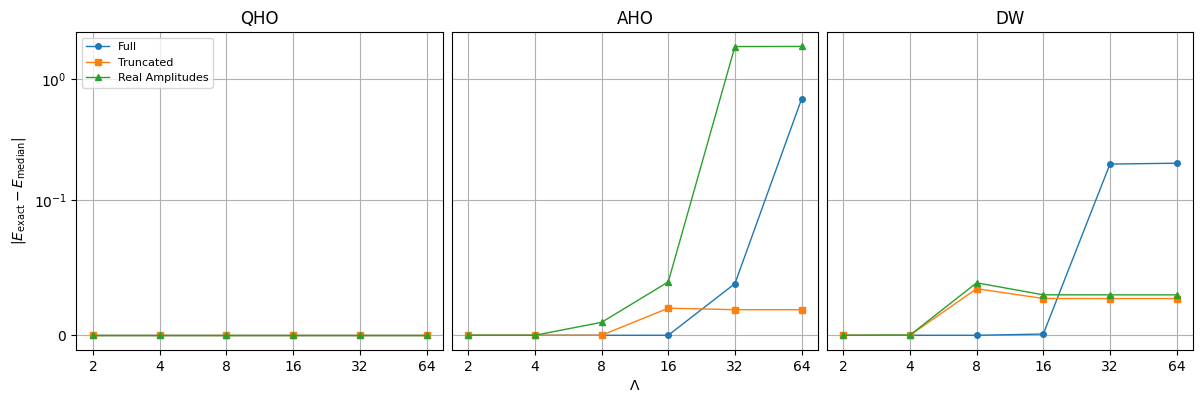}
    \end{subfigure}
    \begin{subfigure}{\linewidth}
        \centering
        \includegraphics[width=0.9\linewidth]{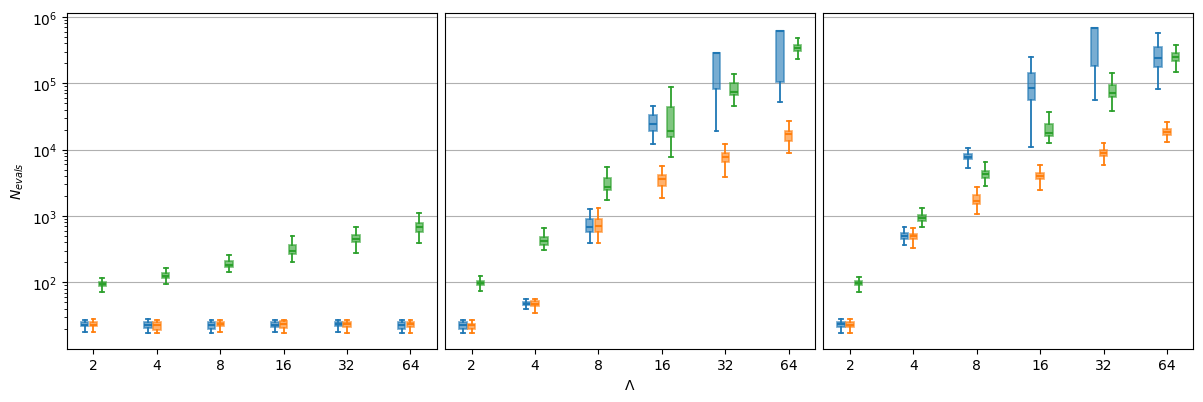}
    \end{subfigure}
    \caption{Comparison of VQE performance between different ans\"atze for 100 independent VQE runs using PennyLane's statevector simulator. `Full' uses the full ansatz returned by the AVQE algorithm while `Truncated' reduces this to at most four gates as shown in \fig{fig:ansatz_circuit_c_limit}. The top plots show the absolute difference between the ground-state energy from exact diagonalization $E_{\text{exact}}$ and the median energy $E_{\text{median}}$ across the 100 independent runs. These points include only runs that converged within a maximum of 10,000 iterations. The bottom plots show the number of quantum circuit evaluations, $N_{\text{evals}}$, aggregated over the 100 VQE runs and including runs that did not converge.}
    \label{fig:ansatze-comparison-None}
\end{figure}

Beginning with statevector simulation, \fig{fig:ansatze-comparison-None} shows the Full ans\"atze outperform both the Truncated and real-amplitudes ans\"atze for $\La \le 16$, as expected. However, for $\La > 16$ we observe the Truncated ans\"atze perform better.
As discussed above, the large number of gates and variational parameters in the Full ans\"atze at larger \La make it hard for the optimizer to correctly converge to the ground state. In fact, in order for any of the AHO runs with $\La \ge 32$ or DW runs with $\La \ge 16$ to converge before reaching a maximum of 10,000 iterations, we had to loosen the tolerance of the optimizer from $10^{-8}$ to $10^{-4}$.

Recall that during the AVQE ansatz construction, the optimal parameters from the previous step are passed forward, which has the advantage of starting each step's VQE in an optimized initial state. When performing a VQE run with the Full ans\"atze outside of the AVQE algorithm, we initialize the ansatz with a random set of parameters. This clearly creates problems for the optimizer and reduces its performance compared to \tab{tab:AVQE-Summary}.  Additionally, for the DW superpotential with $\La \ge 16$ and the AHO superpotential with $\La \ge 32$ the number of variational parameters in the Full ansatz becomes larger than that in the real-amplitudes ansatz. As a result, in many cases, the Full ans\"atze require the largest number of quantum circuit evaluations.  Even when using statevector simulation, the Truncated ans\"atze with only four gates can outperform the other ans\"atze while needing significantly fewer evaluations.

\begin{figure}[htbp]
    \centering
    \begin{subfigure}{\linewidth}
        \centering
        \includegraphics[width=0.9\linewidth]{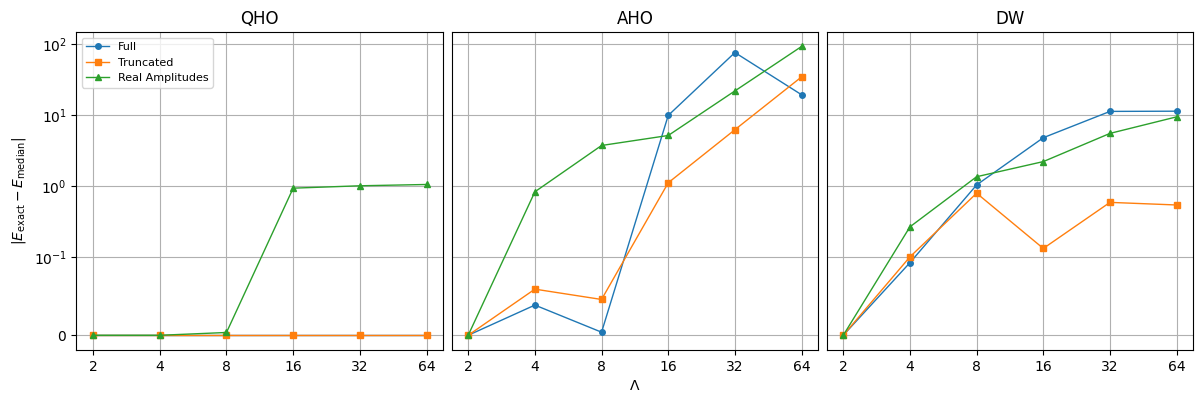}
    \end{subfigure}
    \begin{subfigure}{\linewidth}
        \centering
        \includegraphics[width=0.9\linewidth]{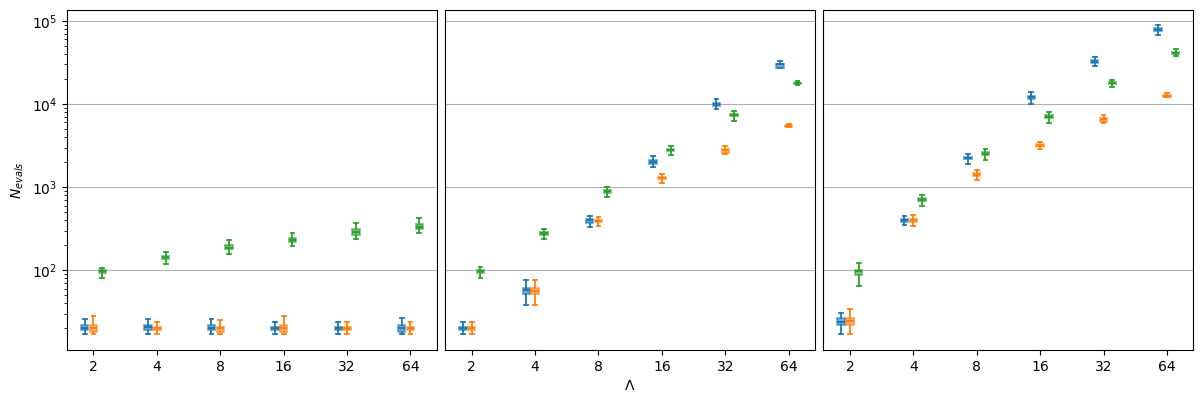}
    \end{subfigure}
    \caption{Comparison of VQE performance between various ans\"atze, as in \fig{fig:ansatze-comparison-None}, but now using PennyLane's default.qubit simulator with 10,000 shots.}
    \label{fig:ansatze-comparison-10k}
\end{figure}

We now repeat these calculations using 10,000 shots rather than statevector simulation (returning the optimizer tolerance to $10^{-8}$).  From the resulting \fig{fig:ansatze-comparison-10k} we can observe that the overall performance with each ansatz declines significantly. Even for the trivial HO superpotential the real-amplitudes ansatz at $\La \ge 8$ struggles to resolve the ground state and appears to converge to the first excited state with $E_1 = 1$. For both the AHO and DW superpotentials, the error in $E_{\text{median}}$ is orders of magnitudes larger than in the statevector case.
However, even with shot noise we can see the benefit of the Truncated ans\"atze at large $\La$.  We can see the Truncated ans\"atze outperform both the Full and real-amplitudes ans\"atze while requiring fewer circuit evaluations. On a NISQ device with gate errors, the performance gap between the Truncated ans\"atze and the others is likely to be even larger. In particular, our Truncated ans\"atze involve only a single two-qubit gate regardless of $\La$, whereas both the full and real-amplitudes ans\"atze involve many more two-qubit gates and larger circuit depths, which would prove problematic on a NISQ device.

\begin{figure}[htbp]
    \centering
    \includegraphics[width=0.9\linewidth]{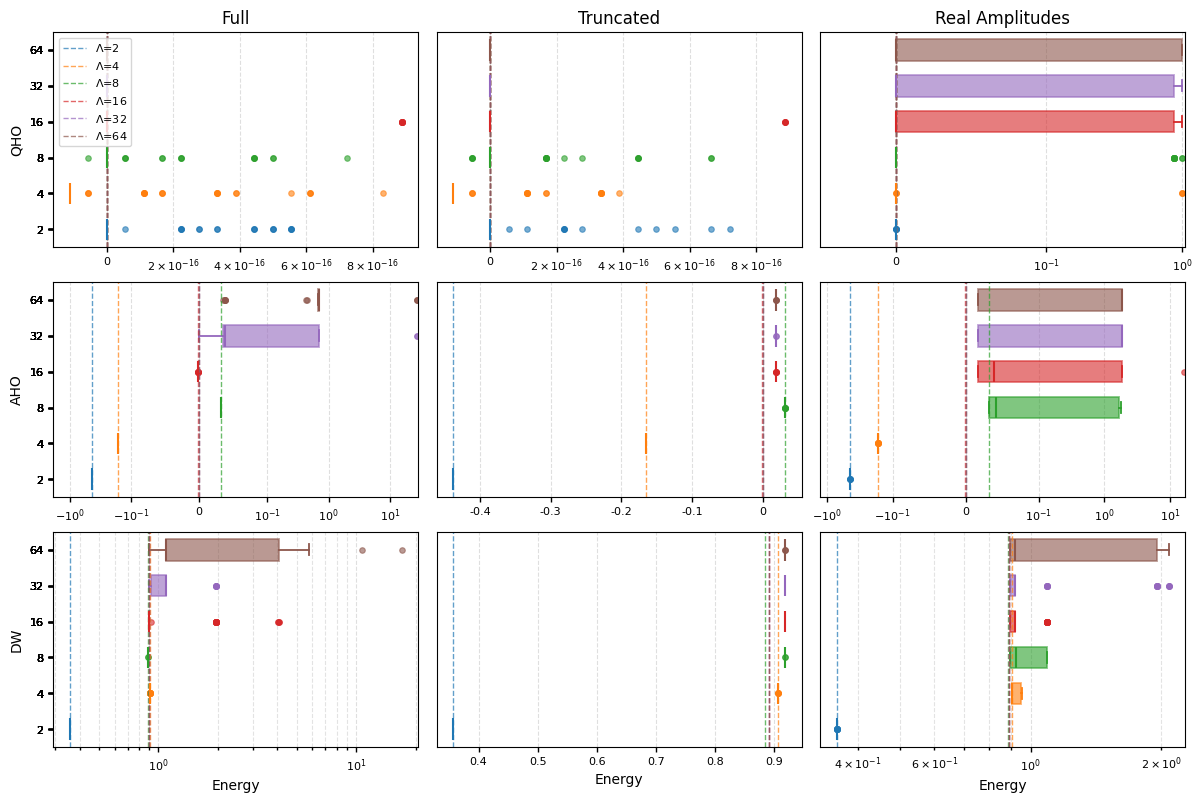}
    \caption{Boxplot distributions for 100 independent VQE runs for each ansatz using statevector simulation.  Each distribution includes only runs that converged within a maximum of 10,000 iterations. Vertical dashed lines indicate the ground-state energy from exact diagonalization, for each value of \La indicated in the legend.}
    \label{fig:boxplot-exact-full-None}
\end{figure}

We can better see the overall distributions of results from the 100 independent VQE runs by employing boxplots, as in Ref.~\cite{Mendicelli:2024ryt}.  Starting with statevector simulation, \fig{fig:boxplot-exact-full-None} shows how creating a problem-tailored ansatz simplifies the energy landscape and reduces the number of local minima: Compared to the more generic real-amplitudes ansatz, the Full and Truncated ans\"atze generally produce much smaller spreads in the energy distributions.  Exceptions to this involve the Full ans\"atze for the AHO and DW superpotentials with $\La \ge 32$.  We attribute this to loosening the tolerance of the optimizer from $10^{-8}$ to $10^{-4}$, which we expect to produce more spread in the VQE energies, especially for systems with a large number of gates.
As expected, for the Truncated ans\"atze with only a maximum of four gates, we see minimal spread. However, we can see that truncating the ans\"atze limits expressivity and for larger \La the VQE energies are unable to reach the exact ground state. Despite this, the distributions for the Truncated ansatz are closer to the exact energies than both the Full and real-amplitudes ans\"atze for $\La \gtrsim  16$.
For the HO superpotential, the real-amplitudes ansatz struggles to reach the ground state consistently and we can see results ranging from the $E = 0$ ground state and the $E_1 = 1$ excited state. When we use the Full or Truncated ans\"atze, which are simply a single RY gate, we can see a consistent convergence to the ground state, with outliers only $\sim$$10^{-16}$ away from the exact energy.

\begin{figure}[htbp]
    \centering
    \includegraphics[width=0.9\linewidth]{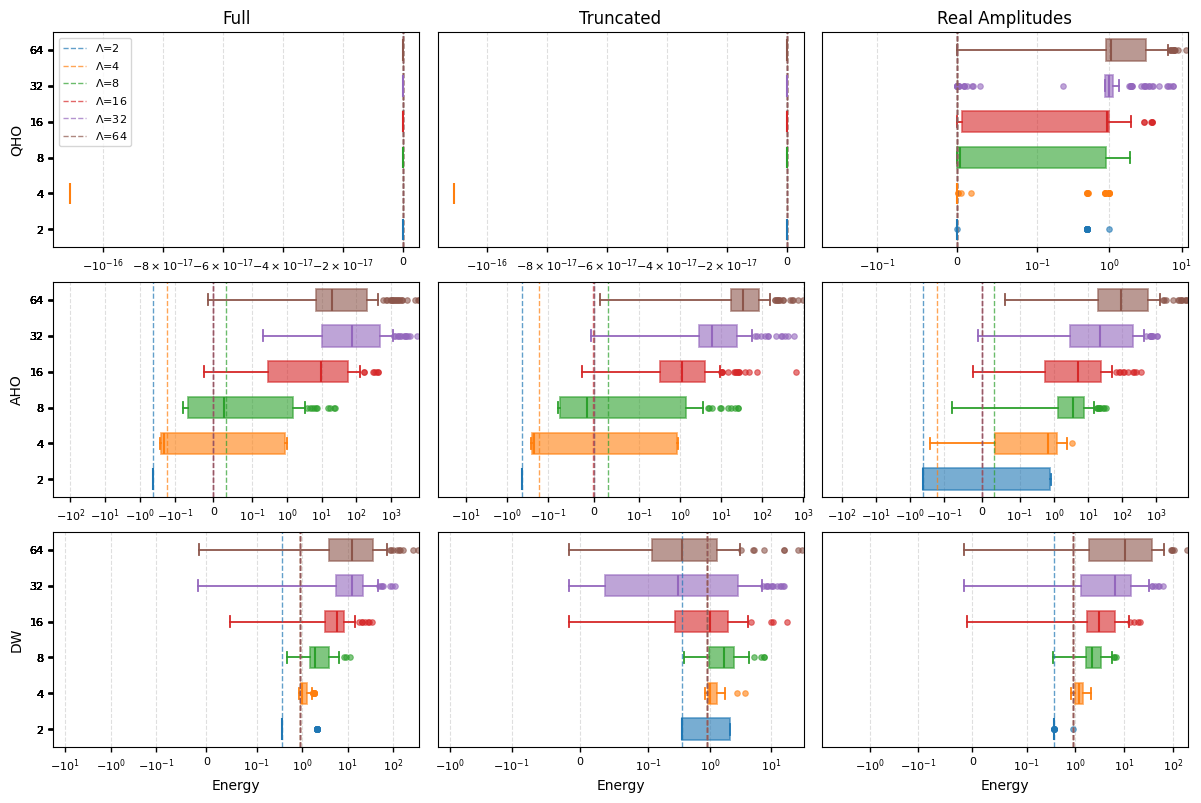}
    \caption{Boxplot distributions for the VQE runs, as in \fig{fig:boxplot-exact-full-None}, but now using 10,000 shots instead of statevector simulation.}
    \label{fig:boxplot-exact-full-10k}
\end{figure}

Finally using 10,000 shots in \fig{fig:boxplot-exact-full-10k}, we see similar Full and Truncated accuracies for the trivial HO case, while the real-amplitudes results spread towards $10$. In general, and as expected, we see significant increases in the spread of results for the AHO and DW superpotentials for all ans\"atze. For the DW superpotential we obtain distributions of energies from $\sim$$0$ up to $\sim$$10^2$ while for the AHO superpotential the range increases to span $\sim$$-1$ up to $\sim$$10^3$. This highlights a key issue with mapping a boson to a qubit device. As discussed in Section~\ref{sec:fock_basis_digitization} and highlighted in \tab{tab:full_H_pauli_strings_energybasis}, as we increase \La for the boson we increase the number of Pauli strings in the digitized Hamiltonian. The number of Pauli strings grows quickly with \La and therefore the cumulative variance resulting from measuring them all is large. This has a big impact on results and is a major reason why we see such a large spread in the distribution. In future work we aim to address this issue by utilizing hybrid systems that can map the bosonic degrees of freedom to continuous-variable `qumodes', which we will discuss in more detail in Section~\ref{sec:conclusions_f_directions}.

\section{Variational quantum deflation}
\label{sec:vqd}

We have seen in \Cref{sec:variational_methods} the challenges introduced by shot noise when estimating the SQM ground-state energy using the VQE algorithm~\cite{Mendicelli:2024ryt}.  This motivates developing alternative analyses of our systems that may be more robust in noisy environments.  In particular, as discussed in \Cref{sec:susy_qm}, we can exploit the pairing structure of the eigenstates to investigate spontaneous supersymmetry breaking.  To do so, we turn to the variational quantum deflation (VQD) algorithm~\cite{Higgott:2018doo} that extends the VQE algorithm to enable us to also estimate excited-state energies.\footnote{In future work it would be interesting to also test the subspace-search VQE~\cite{Nakanishi:2019umy} and compare its performance to the VQD algorithm.} The algorithm achieves this by utilizing the fact that the eigenstates of the system form an orthonormal basis. If two given states are both eigenstates, then their overlap should vanish. Conversely, if the state under consideration begins to approach an eigenstate that has already been resolved, this overlap increases. We can therefore extend the cost function to incorporate the overlap and guide the optimizer to undiscovered eigenstates.
Here is a step-by-step description of estimating the $k$th excited state using the VQD algorithm, which can be compared with the description of the VQE algorithm at the start of \Cref{sec:variational_methods}:
\begin{enumerate}[itemsep=0pt, parsep=0pt]
    \item Prepare a parametrized quantum circuit (ansatz) $\ket{\psi(\boldsymbol{\theta}_k)}$
    \item Measure the expectation value of $\hat{H}$ resulting in the system energy as a function of $\boldsymbol{\theta}_k$
    \[
    E(\boldsymbol{\theta}_k) = \bra{\psi(\boldsymbol{\theta}_k)} \hat{H} \ket{\psi(\boldsymbol{\theta}_k)}.
    \]
    \item Measure the overlap of the current state with each of the $k$ previously discovered states (the ground state and the other $k - 1$ excited states),
    \[
    \lvert \braket{\psi(\boldsymbol{\theta}_k)|\psi(\boldsymbol{\theta}_i)} \rvert^2,
    \]
    where $\boldsymbol{\theta}_i$ are fixed for all $i = 0, \cdots, k - 1$.
    \item Using a classical optimizer, vary $\boldsymbol{\theta}_k$ in order to minimize the extended cost function
    \[
    E(\boldsymbol{\theta}_k) + \sum_{i = 0}^{k - 1} \beta_i \, \lvert \braket{\psi(\boldsymbol{\theta}_k)|\psi(\boldsymbol{\theta}_i)} \rvert^2,
    \]
    where each $\beta_i$ is a real-valued parameter scaling the corresponding overlap penalty term.
    \item Repeat steps 2 to 4 until either the cost function converges or other stopping criteria are met.
\end{enumerate}

We can see the main body of the algorithm remains largely the same as the VQE, and it may face similar challenges with noise.
However, there are reasons to expect to benefit from this approach.
One reason is that it allows us to focus on qualitative features of the energy spectrum rather than the precise value of the ground-state energy.
Specifically, if the VQD algorithm finds a large gap between a single lowest energy and higher energies that are closer together, we can conclude that supersymmetry is preserved.
Similarly, if there is instead a large gap between the third energy level and two lower energies that are closer together, we can conclude that supersymmetry is spontaneously broken.
Any other outcome is inconsistent with the supersymmetry of the Hamiltonian, allowing us to conclude unambiguously that the algorithm has failed and the run simply needs to be discarded.
In addition, while the expectation is that the lowest-energy states will be found first, the VQD algorithm is still able to converge to the ground state later, if excited states are found first --- as in \fig{fig:ansatze-comparison-10k}.
This provides additional opportunities to resolve the ground state every time another state is estimated.
Finally, although we currently use the same ansatz $\psi(\boldsymbol{\theta}_i)$ for each state, this can be generalized to improve the performance of the algorithm.

Compared to the VQE, the VQD algorithm includes the additional step of calculating the overlap of the current ansatz state with each of the eigenstates found previously. It is important to note that when using a real device we do not have access to the statevector and must estimate the overlap using a quantum circuit. The SWAP test circuit is a popular choice and is often used in quantum error correction to calculate the overlap between two states without directly measuring and collapsing the original state~\cite{SWAP:barenco1996,SWAP:Foulds_2021,SWAP:Buhrman_2001}. The need for indirect measurement requires an ancillary qubit in the SWAP test circuit, resulting in a total of $2n+1$ qubits for the overall circuit.

Since maintaining the original state is not required for our problem, we can instead employ the destructive SWAP (DSWAP) test~\cite{DSWAP:PhysRevA.87.052330}. The DSWAP test has many advantages over the standard SWAP test. First, it doesn't require an ancillary qubit, reducing the total number of qubits to $2n$. In addition, the circuit complexity is drastically reduced for the DSWAP test. The SWAP test circuit utilizes the three-qubit CSWAP gate which is decomposed into multiple CNOT, Toffoli and single-qubit gates in order to be executed on a real device~\cite{CSWAP:M_Q_Cruz_2024}. In comparison, the DSWAP circuit utilizes the Hadamard and CNOT gates which are native to most quantum devices, resulting in improved performance and significantly shallower circuit depth. This makes the DSWAP circuit preferable for computing the overlaps for the VQD algorithm on NISQ devices.

In principle we could estimate all eigenstates of a system with this method.  In practice, more eigenstates require more computationally intensive overlap computations and make it more challenging fro the optimizer to converge.  We therefore search for only the ground state and the first two excited states. From these first three energy levels we can construct the ratio
\begin{equation}\label{eq:vqd-ratio}
    R = \frac{E_2 - E_1}{E_2 - E_0}
\end{equation}
to quantify the pairing structure of our system. Preserved supersymmetry requires a single $E_0=0$ ground state, with all subsequent energy levels paired so that $E_1= E_2 > 0$, corresponding to $R=0$. The case of spontaneous supersymmetry breaking instead requires a degenerate pair of ground states with positive energies $E_0= E_1 > 0$, and therefore $R=1$. \Cref{app:energyspectrum_artifacts} includes examples of both of these cases from exact diagonalization calculations for small $\La$.

As was the case for the VQE, choosing a suitable optimizer and ansatz is important for the VQD algorithm.  We continue to use the COBYQA optimizer for our VQD analyses.  Unfortunately the efficient problem-tailored ground-state ans\"atze we were able to find through the AVQE algorithm in \Cref{ansatz_extrapolation} are not sufficient for the VQD. Since we are now looking for excited states as well as the ground state, we need more general ans\"atze than those shown in \fig{fig:ansatz_circuit_c_limit}.  While future work will explore generalizations of the AVQE algorithm to also find efficient excited-state ans\"atze, for now we revert to using a real-amplitudes ansatz.

\subsection{Choosing $\beta$}
Before performing VQD analyses it is important for us to choose suitable values for the parameters $\beta_i$ that scale the overlap penalty terms in the extended cost function.
For simplicity we take all of these parameters to be the same $\beta$.
Qualitatively, if $\beta$ is too small then the penalty in the cost function may not be large enough to guide the optimizer away from the overlapping state. If $\beta$ is too large then the penalty may dominate the cost function and make the optimizer insensitive to the energy itself.
Ref.~\cite{Higgott:2018doo} argues that $\beta$ should be larger than all energy gaps between the states under consideration.
In \tab{tab:VQD-energygaps} we collect the energy gaps between the first three energy levels for each superpotential, from exact diagonalization for small $\La \leq 16$.  (The gap $\Delta E_{02} = \Delta E_{01} + \Delta E_{12}$.)  In all cases except the AHO superpotential with the severe truncation $\La=2$, the energy gaps are $\Delta E < 2$, suggesting we should consider $\beta \gtrsim 2$.

\begin{table}[htbp]
\centering
\begin{minipage}[t]{0.32\textwidth}
\centering
    \begin{tabular}{lrrr}
    \hline
    $W(\qhat)$ & \La & $\Delta E_{01}$ & $\Delta E_{12}$ \\
    \hline
    HO & 2 & 0.000 & 1.000 \\
    HO & 4 & 1.000 & 0.000 \\
    HO & 8 & 1.000 & 0.000 \\
    HO & 16 & 1.000 & 0.000 \\
    \hline
    \end{tabular}
\end{minipage}
\hfill
\begin{minipage}[t]{0.32\textwidth}
\centering
    \begin{tabular}{lrrr}
    \hline
    $W(\qhat)$ & \La & $\Delta E_{01}$ & $\Delta E_{12}$ \\
    \hline
    AHO & 2 & 0.000 & 2.500 \\
    AHO & 4 & 0.838 & 0.995 \\
    AHO & 8 & 1.648 & 0.153 \\
    AHO & 16 & 1.679 & 0.009 \\
    \hline
    \end{tabular}
\end{minipage}
\hfill
\begin{minipage}[t]{0.32\textwidth}
\centering
    \begin{tabular}{lrrr}
    \hline
    $W(\qhat)$ & \La & $\Delta E_{01}$ & $\Delta E_{12}$ \\
    \hline
    DW & 2 & 0.414 & 0.707 \\
    DW & 4 & 0.044 & 0.745 \\
    DW & 8 & 0.003 & 1.806 \\
    DW & 16 & 0.000 & 1.842 \\
    \hline
    \end{tabular}
\end{minipage}
\caption{Energy gaps $\Delta E_{01}$ and $\Delta E_{12}$ for each superpotential, from exact diagonalization with small $\La \leq 16$.}
\label{tab:VQD-energygaps}
\end{table}

In \tab{tab:vqd-betas} we test this expectation by varying $\beta$ from 0.5 to 20 in VQD calculations using statevector simulation for each superpotential with $\La = 16$. As expected, for $\beta \le 2$ the VQD produces a ratio $R$ increasing different than that from exact diagonalization.  In particular, for the HO and AHO superpotentials these small $\beta$ can lead to $R\approx0.5$, inconsistent with both preserved and spontaneously broken supersymmetry.  The AHO case also exhibits a similar issue for large $\beta = 20$.  Based on these results we choose $\beta=5$ for our VQD computations with other \La reported in the next subsection.  While this choice is sufficient for the current work, in the future we may refine this by setting separate $\beta_i$ values for each superpotential and overlap term.

\begin{table}[htbp]
  \centering
  \begin{subtable}{0.48\textwidth}
    \centering
    \caption*{HO}
    \resizebox{\textwidth}{!}{
      \begin{tabular}{cccccc}
      \toprule
      $\beta$ & Converged Runs & $N_{\text{iters}}$ & $R_{\text{exact}}$ & $R_{\text{med}}$ & $\Delta R$ \\
      \midrule
      0.5  & 100/100 & 1002 & 0.0000 & 0.5000 & 0.5000 \\
      1.0  & 100/100 &  999 & 0.0000 & 0.0000 & 0.0000 \\
      2.0  & 100/100 & 1000 & 0.0000 & 0.0000 & 0.0000 \\
      3.0  & 100/100 &  988 & 0.0000 & 0.0000 & 0.0000 \\
      4.0  & 100/100 &  941 & 0.0000 & 0.0000 & 0.0000 \\
      5.0  & 100/100 &  945 & 0.0000 & 0.0000 & 0.0000 \\
      10.0 & 100/100 &  972 & 0.0000 & 0.2265 & 0.2265 \\
      20.0 & 100/100 & 1044 & 0.0000 & 0.0000 & 0.0000 \\
      \bottomrule
      \end{tabular}
    }
  \end{subtable}\hfill
  \begin{subtable}{0.48\textwidth}
    \centering
    \caption*{AHO}
    \resizebox{\textwidth}{!}{
      \begin{tabular}{cccccc}
      \toprule
      $\beta$ & Converged Runs & $N_{\text{iters}}$ & $R_{\text{exact}}$ & $R_{\text{med}}$ & $\Delta R$ \\
      \midrule
      0.5  & 91/100  & 5808 & 0.0053 & 0.5000 & 0.4947 \\
      1.0  & 92/100  & 5987 & 0.0053 & 0.4016 & 0.3964 \\
      2.0  & 94/100  & 5808 & 0.0053 & 0.1644 & 0.1591 \\
      3.0  & 96/100  & 5030 & 0.0053 & 0.0425 & 0.0372 \\
      4.0  & 96/100  & 5249 & 0.0053 & 0.0926 & 0.0873 \\
      5.0  & 96/100  & 4993 & 0.0053 & 0.0676 & 0.0623 \\
      10.0 & 98/100  & 4456 & 0.0053 & 0.0921 & 0.0868 \\
      20.0 & 96/100  & 4242 & 0.0053 & 0.6979 & 0.6926 \\
      \bottomrule
      \end{tabular}
    }
  \end{subtable}

  \vspace{1em}

  \centering
  \begin{subtable}{0.48\textwidth}
    \centering
    \caption*{DW}
    \resizebox{\textwidth}{!}{
      \begin{tabular}{cccccc}
      \toprule
      $\beta$ & Converged Runs & $N_{\text{iters}}$ & $R_{\text{exact}}$ & $R_{\text{med}}$ & $\Delta R$ \\
      \midrule
      0.5  & 99/100  & 3209 & 1.0000 & 0.9595 & 0.0405 \\
      1.0  & 100/100 & 2724 & 1.0000 & 0.9792 & 0.0207 \\
      2.0  & 90/100  & 3850 & 1.0000 & 0.9894 & 0.0105 \\
      3.0  & 100/100 & 2219 & 1.0000 & 0.9896 & 0.0104 \\
      4.0  & 100/100 & 2137 & 1.0000 & 0.9897 & 0.0103 \\
      5.0  & 100/100 & 2256 & 1.0000 & 0.9897 & 0.0103 \\
      10.0 & 100/100 & 1982 & 1.0000 & 0.9906 & 0.0094 \\
      20.0 & 100/100 & 1945 & 1.0000 & 0.9905 & 0.0095 \\
      \bottomrule
      \end{tabular}
    }
  \end{subtable}
  \caption{Comparison of 100 independent VQD runs for each superpotential and $\La=16$ using statevector simulation and varying values of $\beta$. $N_{\text{iters}}$ is the mean number of iterations needed for the VQD to resolve the the energy levels and includes data from runs that did not converge within the maximum of 10,000 iterations per state. The median ratio $R_{\text{med}}$ is calculated from \eq{eq:vqd-ratio} using converged runs only, while $R_{\text{exact}}$ comes from exact diagonalization and $\Delta R$ is the different between the two.}
  \label{tab:vqd-betas}
\end{table}

\subsection{Results}
\begin{figure}[htbp]
    \centering
    \includegraphics[width=0.9\textwidth]{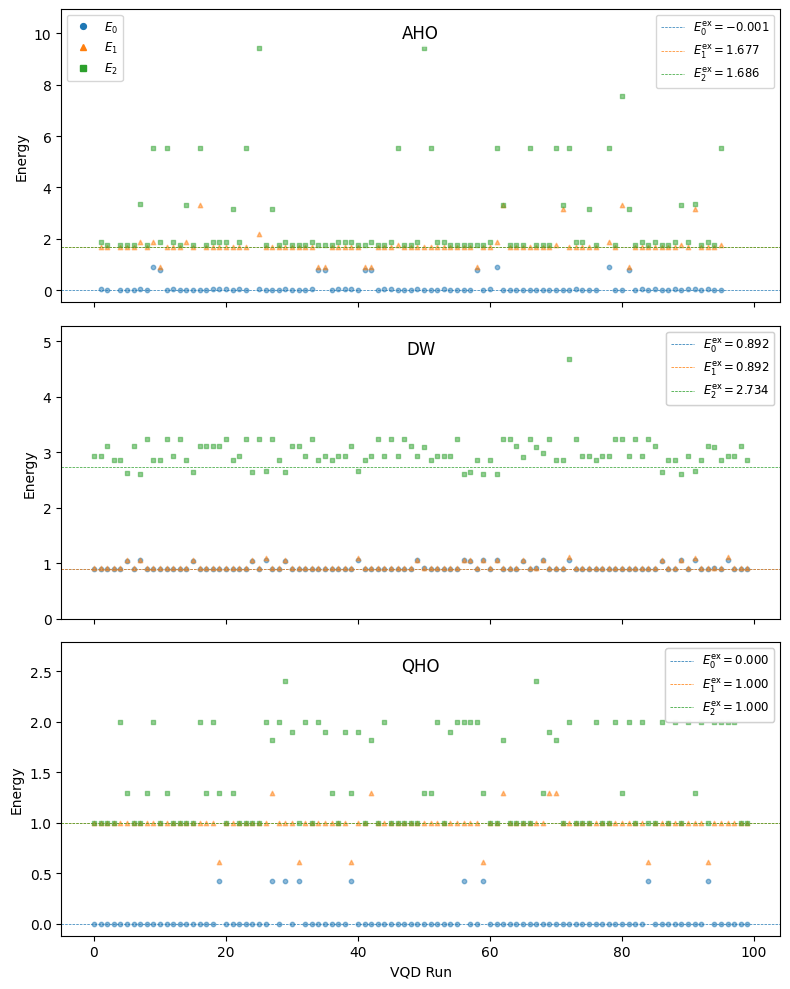}
    \caption{Results for 100 independent VQD runs using $\La=16$, $\beta=5$ and statevector simulation. Only runs where all three energy levels converge are shown. Plotted points are the individual energies returned from each VQD run and dashed lines indicate the true energies $E_i^{\text{ex}}$ from exact diagonalization (listed in each legend).}
    \label{fig:VQD-Single16}
\end{figure}

\Cref{fig:VQD-Single16} shows results for 100 independent VQD runs using statevector simulation for each superpotential with $\La = 16$. Despite there being fluctuations to higher excited states and pseudo-levels, we can clearly see the expected spacing between these three energy levels. The true energies from exact diagonalization are shown as the dashed lines in the plot and the numerical values are listed in each legend. As expected, for the HO and AHO superpotentials, we can see a single $E_0=0$ ground state and $E_1\approx E_2 > 0$. For the DW superpotential we see that larger fluctuations in $E_2$ don't disguise the clear gap between $E_2$ and $E_0\approx E_1 > 0$, confirming spontaneous supersymmetry breaking.

\tab{tab:VQD-Summary-SV} summarizes similar VQD analyses for $\La$ up to 32. The four right-most columns are computed from exact diagonalization and reveal the effect on $R$ of truncating the bosonic modes.  For small \La both the HO and AHO cases produce $R=1$, erroneously indicating spontaneous supersymmetry breaking. By increasing $\La$, we reduce the explicit supersymmetry breaking from this truncation, and the ratio approaches the correct value $R=0$. Similarly, for the DW case we have an ambiguous $R = 0.631$ for $\La = 2$, but $R$ approaches the correct value $R=1$ as \La increases.

The need for a more complex cost function means that we suffer from reduced precision in the ground-state energies compared to the VQE. However, this is not problematic so long as the spacing between the energy levels suffices to distinguish between the supersymmetric vs.\ spontaneously broken pairing structures. In \tab{tab:VQD-Summary-SV} we can clearly see $R$ approaching the correct value for each superpotential as \La increases. The ability to extract this information from less-precise energies is promising in the context of NISQ devices where robustness to noise is important.

Since, in essence, we are performing three VQE calculations in each VQD run, significantly more quantum resources are required to perform the latter.  We can note in \tab{tab:VQD-Summary-SV} that even the trivial HO superpotential with $\La=2$ requires an average of 237 iterations per run (summed over all three energy levels). This number can grow quickly, and exceeds 10,000 for the AHO superpotential with $\La=32$, highlighting how costly this algorithm can become.

\begin{table}[htbp]
\centering
\resizebox{0.9\textwidth}{!}{%
\begin{tabular}{lrccc|cccc|cccc|}
\cline{6-9} \cline{10-13}
\multicolumn{5}{l|}{} & \multicolumn{4}{c|}{VQD} & \multicolumn{4}{c|}{Exact Diagonalization} \\
\hline
$W(\qhat)$   &   \La &   Converged Runs & $N_{\text{iters}}$ & $N_{\text{evals}}$ & $E_{0}$ &     $E_{1}$ &    $E_{2}$ &   $\text{R}$ &   $E_{0}$ &   $E_{1}$ &   $E_{2}$ &   $\text{R}$  \\
\hline
    HO & 2 & 100/100 & 237 & 609 & 0.0000 & 0.0000 & 1.0000 & 1.0000 & 0.0000 & 0.0000 & 1.0000 & 1.0000 \\
    HO & 4 & 100/100 & 352 & 938 & 0.0000 & 1.0000 & 1.0000 & 0.0000 & 0.0000 & 1.0000 & 1.0000 & 0.0000 \\
    HO & 8 & 100/100 & 468 & 1228 & 0.0000 & 1.0000 & 1.0000 & 0.0000 & 0.0000 & 1.0000 & 1.0000 & 0.0000 \\
    HO & 16 & 100/100 & 721 & 1963 & 0.0000 & 1.0000 & 1.2929 & 0.2265 & 0.0000 & 1.0000 & 1.0000 & 0.0000 \\
    HO & 32 & 100/100 & 1040 & 2901 & 0.0000 & 1.0000 & 1.0000 & 0.0000 & 0.0000 & 1.0000 & 1.0000 & 0.0000 \\
 \hline
 \\
 \hline
    AHO & 2 & 100/100 & 231 & 586 & -0.4375 & -0.4375 & 2.0625 & 1.0000 & -0.4375 & -0.4375 & 2.0625 & 1.0000 \\
    AHO & 4 & 100/100 & 464 & 1796 & -0.1648 & 0.6733 & 1.6679 & 0.5427 & -0.1648 & 0.6733 & 1.6679 & 0.5427 \\
    AHO & 8 & 100/100 & 1461 & 12000 & 0.0320 & 1.6803 & 1.8696 & 0.1030 & 0.0320 & 1.6802 & 1.8335 & 0.0851 \\
    AHO & 16 & 96/100 & 4551 & 83140 & 0.0172 & 1.6884 & 1.8589 & 0.0926 & -0.0012 & 1.6775 & 1.6864 & 0.0053 \\
    AHO & 32 & 64/100 & 10061 & 388260 & 0.0174 & 1.6883 & 1.7609 & 0.0416 & 0.0000 & 1.6865 & 1.6866 & 0.0000 \\
 \hline
 \\
 \hline
    DW & 2 & 100/100 & 261 & 604 & 0.3572 & 0.7714 & 1.4786 & 0.6306 & 0.3572 & 0.7714 & 1.4786 & 0.6306 \\
    DW & 4 & 100/100 & 456 & 3522 & 0.9066 & 0.9506 & 1.6957 & 0.9441 & 0.9066 & 0.9506 & 1.6957 & 0.9441 \\
    DW & 8 & 100/100 & 889 & 17758 & 0.8934 & 0.9172 & 2.9101 & 0.9882 & 0.8846 & 0.8877 & 2.6939 & 0.9983 \\
    DW & 16 & 100/100 & 2183 & 95808 & 0.8937 & 0.9149 & 2.9416 & 0.9897 & 0.8916 & 0.8916 & 2.7341 & 1.0000 \\
    DW & 32 & 95/100 & 4090 & 374909 & 0.8937 & 0.9148 & 2.9416 & 0.9897 & 0.8916 & 0.8916 & 2.7340 & 1.0000 \\
\hline
\end{tabular}
}
 \caption{Results for 100 independent VQD runs using $\beta=5$, statevector simulation and a real-amplitudes ansatz. $N_{\text{iters}}$ and $N_{\text{evals}}$ are the mean numbers of iterations and evaluations needed for the VQD to resolve all three energy levels, and include data from runs that did not converge within the maximum of 10,000 iterations per state.  The final four columns come from exact diagonalization, for comparison.}
 \label{tab:VQD-Summary-SV}
\end{table}

When running the VQD with shot noise it is important to consider the statistical error from calculating each overlap using the DSWAP circuit with a finite number of shots. If the error in this calculation of the overlap is large, it will also impact the overall cost function evaluation. It is important in general to be conscious of the quantum resources we are utilizing. The overlap test requires a separate circuit evaluation with its own number of shots. At the second excited state the overlap test is computed twice meaning for each iteration of the VQD there are three additional quantum circuit evaluations, each with shot noise. One can appreciate this resource burden will likely prove to be an obstacle to running this algorithm on a currently available quantum device. \tab{tab:VQD-Summary-10k} summarizes results using 10,000 shots for both the energy evaluation and DSWAP test. Results are poor and we are unable to clearly resolve the ratios we would expect for each superpotential.
Similar to the VQE case, the large statistical fluctuations from measuring with a finite number of shots cause problems for the optimizer, which often converges to the wrong energy levels.

While our SQM results for the VQD algorithm using statevector simulation are promising, the more realistic approach of including shot noise indicates that more work is needed to make this method viable on existing quantum devices.  This is further motivation for future work to avoid the rapid growth in the number of Pauli strings shown in \fig{fig:fits_pauli_fock_basis}, for instance by utilizing hybrid systems that can map the bosonic degrees of freedom to continuous-variable qumodes.

\begin{table}[htbp]
\centering
\resizebox{0.9\textwidth}{!}{%
\begin{tabular}{lrccc|cccc|cccc|}
\cline{6-9} \cline{10-13}
\multicolumn{5}{l|}{} & \multicolumn{4}{c|}{VQD} & \multicolumn{4}{c|}{Exact Diagonalization} \\
\hline
$W(\qhat)$   &   \La &   Converged Runs & $N_{\text{iters}}$ & $N_{\text{evals}}$ & $E_0$ &     $E_1$ &    $E_2$ &   $R$ &   $E_0$ &   $E_1$ &   $E_2$ &   $R$  \\
\hline
    HO & 2 & 100/100 & 163 & 549 & -0.0967 & 0.0000 & 0.8570 & 0.8986 & 0.0000 & 0.0000 & 1.0000 & 1.0000 \\
    HO & 4 & 100/100 & 205 & 778 & -0.0000 & 0.8586 & 0.9085 & 0.0550 & 0.0000 & 1.0000 & 1.0000 & 0.0000 \\
    HO & 8 & 100/100 & 256 & 1026 & 0.0000 & 0.8923 & 1.8443 & 0.5162 & 0.0000 & 1.0000 & 1.0000 & 0.0000 \\
    HO & 16 & 100/100 & 333 & 1335 & 0.0000 & 0.9710 & 1.9014 & 0.4893 & 0.0000 & 1.0000 & 1.0000 & 0.0000 \\
    HO & 32 & 100/100 & 398 & 1618 & 0.0204 & 1.0630 & 2.8190 & 0.6274 & 0.0000 & 1.0000 & 1.0000 & 0.0000 \\
 \hline
 \\
 \hline
    AHO & 2 & 100/100 & 166 & 553 & -0.5390 & -0.4375 & 1.9025 & 0.9584 & -0.4375 & -0.4375 & 2.0625 & 1.0000 \\
    AHO & 4 & 100/100 & 237 & 1328 & -0.2273 & 1.3510 & 2.4215 & 0.4041 & -0.1648 & 0.6733 & 1.6679 & 0.5427 \\
    AHO & 8 & 100/100 & 268 & 3244 & 0.0030 & 1.5359 & 5.0479 & 0.6962 & 0.0320 & 1.6802 & 1.8335 & 0.0851 \\
    AHO & 16 & 100/100 & 310 & 9105 & 0.1235 & 1.9438 & 5.8245 & 0.6807 & -0.0012 & 1.6775 & 1.6864 & 0.0053 \\
    AHO & 32 & 100/100 & 362 & 23094 & 0.8468 & 3.9178 & 10.0467 & 0.6662 & 0.0000 & 1.6865 & 1.6866 & 0.0000 \\
 \hline
 \\
 \hline
    DW & 2 & 100/100 & 190 & 627 & 0.3572 & 0.7714 & 1.3368 & 0.5771 & 0.3572 & 0.7714 & 1.4786 & 0.6306 \\
    DW & 4 & 100/100 & 235 & 2656 & 0.8940 & 1.0537 & 2.0118 & 0.8571 & 0.9066 & 0.9506 & 1.6957 & 0.9441 \\
    DW & 8 & 100/100 & 280 & 8347 & 0.6628 & 1.9379 & 3.9414 & 0.6111 & 0.8846 & 0.8877 & 2.6939 & 0.9983 \\
    DW & 16 & 100/100 & 322 & 21898 & 0.2953 & 2.5634 & 5.6614 & 0.5773 & 0.8916 & 0.8916 & 2.7341 & 1.0000 \\
    DW & 32 & 100/100 & 366 & 53205 & 0.0682 & 4.2688 & 7.2665 & 0.4164 & 0.8916 & 0.8916 & 2.7340 & 1.0000 \\
\hline
\end{tabular}
}
 \caption{Results for 100 independent VQD runs, as in \cref{tab:VQD-Summary-SV}, but now using 10,000 shots instead of statevector simulation for both the energy evaluation and each DSWAP test.}
 \label{tab:VQD-Summary-10k}
\end{table}

\section{Conclusions and future directions}
\label{sec:conclusions_f_directions}

Motivated by the rapid and ongoing development of quantum computing, in this work we explored the application of quantum variational methods to investigate a minimal supersymmetric model --- a single-site interacting fermion--boson system. This system is simple enough to be within reach of existing quantum devices, while still exhibiting spontaneous supersymmetry breaking which is generically challenge to analyze due to a severe sign problem in standard Monte Carlo calculations.  The presence of interacting bosonic and fermionic degrees of freedom is another significant aspect of the system, requiring supersymmetry-breaking truncation of the bosonic degree of freedom to retain only $\La = 2^B$ modes.  The implications of supersymmetry for the energy spectrum enable us to study spontaneous supersymmetry breaking simply by estimating the ground state energy using the VQE algorithm, or by examining the pairing structure of the lowest excited states using the VQD algorithm.

Both algorithms rely on a classical optimizer, and while we find that differential evolution converges more reliably to the ground state it requires a large number of circuit evaluations to do so. Since we are targeting NISQ devices we adopted the COBYQA optimizer to achieve the best balance between performance and quantum resource requirements.  Both algorithms also rely on an ansatz for the targeted state(s).  Because generic ans\"atze involve large numbers of gates and variational parameters, we implemented AVQE ansatz construction in order to systematically construct efficient problem-tailored ans\"atze.  In particular, we were able to propose suitable VQE ans\"atze for large systems, involving at most four variational parameters, by extrapolating those obtained for smaller systems.  In the presence of shot noise, these `Truncated' ans\"atze outperformed other options while requiring fewer quantum circuit evaluations.  However, the rapid increase in the number of Pauli strings for larger boson truncations \La makes it challenging to obtain precise results even with these ans\"atze.  The construction of efficient problem-tailored ans\"atze and analysis of the role of shot noise are key new features of this work compared to earlier Refs.~\cite{Culver:2021rxo, Culver:2023iif, Schaich:2024bmg, Mendicelli:2024ryt}.

Finally we explored the VQD algorithm as an alternative approach to assessing spontaneous supersymmetry breaking that may be more robust to noise. While results using statevector simulation were promising, the more realistic approach of including shot noise again made it challenging to distinguish between supersymmetric and spontaneously broken systems.  Additionally, the greater quantum resource requirements of the VQD compared to the VQE introduces further challenges to using this algorithm on a NISQ device.

There are many future directions in which this research can now be taken.
We have already mentioned the possibility of comparing the VQE and VQD against other noise-resilient methods including quantum imaginary time evolution~\cite{Motta:2019yya}, the quantum alternating operator ansatz~\cite{Farhi:2014ych, Hadfield:2017yqz}, and the subspace-search VQE~\cite{Nakanishi:2019umy}.
Pursuing the generalization of the AVQE ansatz construction to also find efficient excited-state ans\"atze could significantly improve the feasibility of the VQD algorithm.
Even more ambitiously, we can explore the use of hybrid systems involving coupled qubits for fermions and continuous-variable qumodes for bosons~\cite{Liu:2024mbr, Crane:2024tlj, Araz:2024dcy}.
This promises the significant advantage of using a bosonic system to encode bosonic degrees of freedom, avoiding the need to spread out the truncated bosons across a large number of qubits, which contributes to the rapid growth in the number of Pauli strings as \La increases.

Finally, the methods we develop in the context of supersymmetric quantum mechanics can then be applied to explore more computationally demanding and physically rich fermion--boson systems.
These include $(0+1)$-dimensional supersymmetric matrix models~\cite{Buividovich:2022jgv, Jha:2024rxz} as well as quantum field theories in higher dimensions such as the $(1+1)$-dimensional $\mathcal N = 1$ Wess--Zumino model which features truly dynamical supersymmetry breaking.
Initial explorations of using the VQE and VQD algorithms for the Wess--Zumino model have already appeared~\cite{Culver:2023iif, Schaich:2024bmg}, and there is promising scope to improve upon that work by applying the AVQE ansatz construction.
This will not be trivial, and we expect initial steps will involve carefully reviewing the structure of the Hamiltonian and exactly diagonalizing small systems to gain insight into suitable initial basis states to use as a starting point.
As our research advances, we can also look forward to generalizing beyond the Wess--Zumino model and investigating more complicated $(1+1)$-dimensional supersymmetric quantum field theories, including super-Yang--Mills~\cite{Catterall:2017xox}, super-QCD~\cite{Catterall:2015tta} and the supersymmetric Gross--Neveu--Yukawa model~\cite{Fitzpatrick:2019cif}.

\vspace{8 pt}
\noindent \textbf{\textsc{Acknowledgments:}} We thank Chris Culver for previous collaboration on Refs.~\cite{Culver:2021rxo, Culver:2023iif, Schaich:2024bmg}.
Numerical calculations were carried out at the University of Liverpool.
JK was supported through the Liv.Inno Centre for Doctoral Training, funded by the UK's Science and Technology Facilities Council (STFC) under grant agreement {ST/W006766/1}.
EM and DS were supported by UK Research and Innovation Future Leader Fellowship {MR/X015157/1}, with additional support for DS from STFC consolidated grant {ST/X000699/1}.


\appendix

\section{The energy spectrum and the effects of bosonic truncation}\label{app:energyspectrum_artifacts}

In this appendix, we use exact diagonalization to present the low-lying energy spectrum of the theory for each of the three superpotentials under consideration.
Recall that we employ the Fock basis representation for the bosonic operators \phat and $\qhat$, with a truncation that allows only \La bosonic modes.  We begin by analyzing the case where all dimensionless parameters are set to unity, $m=g=\mu=1$, and the truncation \La is increased. We then examine how variations in these parameters affect the energies, confirming that truncation artifacts decrease as the number of included bosonic modes increases. Recall as well that supersymmetry requires a single zero-energy ground state, with all excited states appearing in pairs, while spontaneous supersymmetry breaking corresponds to a degenerate pair of ground states with positive energy.

\begin{figure}[htbp]
    \centering
    \begin{subfigure}{0.32\textwidth}
        \centering
        \includegraphics[width=\linewidth]{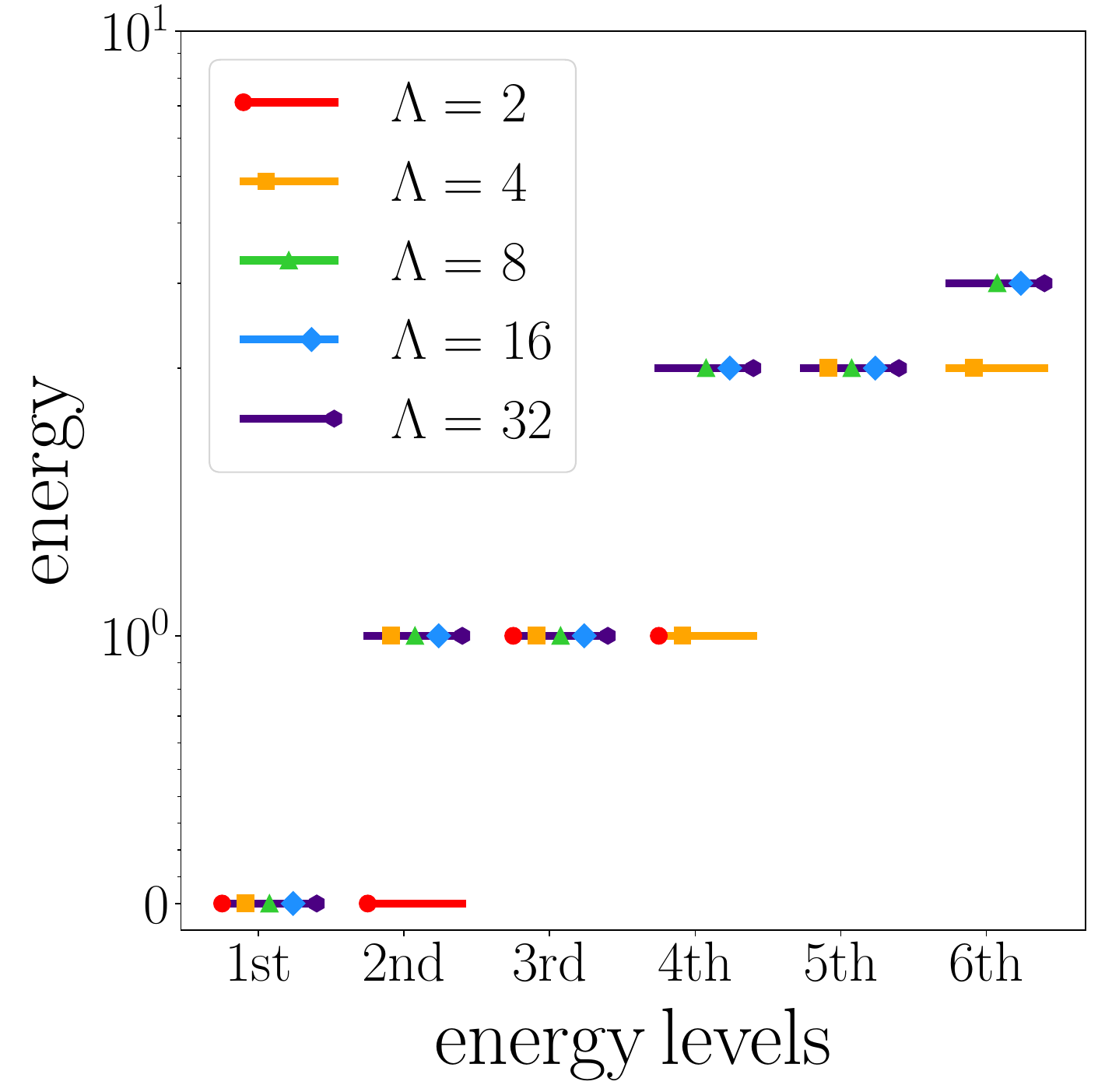}
        {Harmonic Oscillator}
    \end{subfigure}
    \hfill
    \begin{subfigure}{0.32\textwidth}
        \centering
        \includegraphics[width=\linewidth]{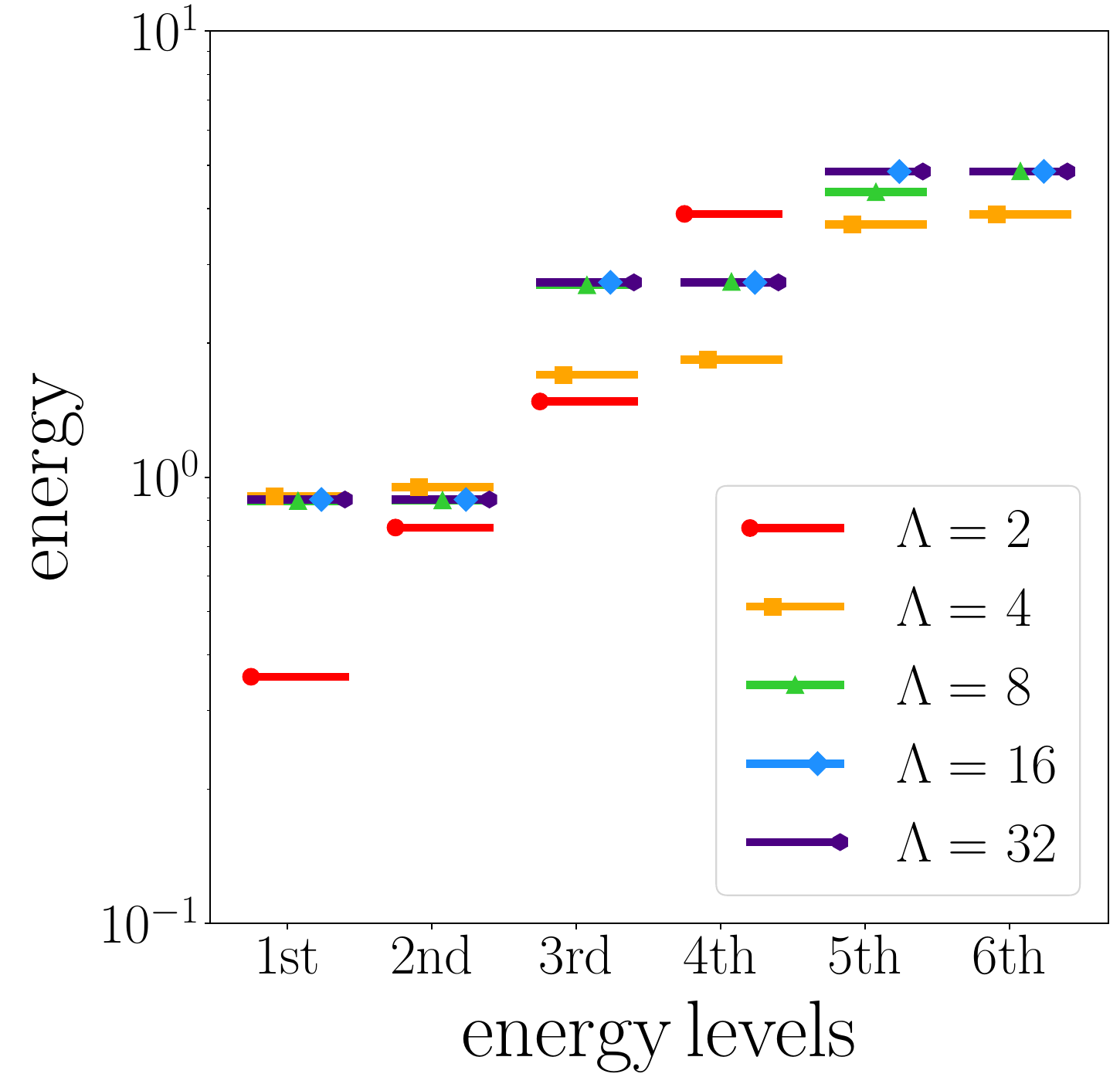}
        {Double well}
    \end{subfigure}
    \hfill
    \begin{subfigure}{0.32\textwidth}
        \centering
    \includegraphics[width=\linewidth]{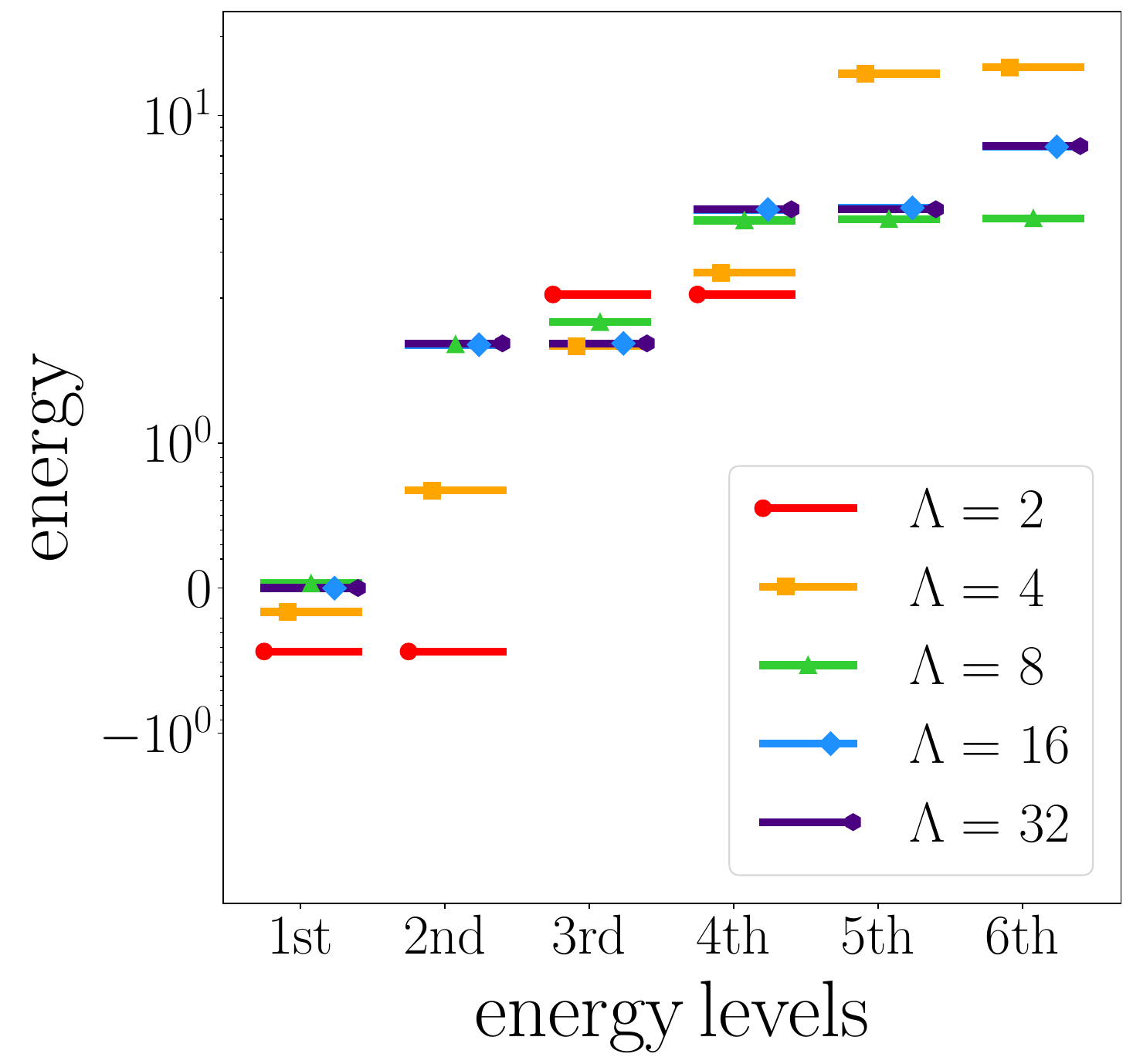}
    {Anharmonic Oscillator}
    \end{subfigure}
    \caption{The low-lying energy spectrum of SQM is shown for each superpotential, with $m=g=\mu=1$. The first six energy levels are displayed for increasing values of the truncation parameter $\La$. Each energy level is represented by a line, with different colors corresponding to different $\La$. To distinguish overlapping energy levels from different \La values, distinct markers are used for each truncation. The plots demonstrate rapid convergence of the spectrum as \La increases, specifically for $\La > 2$ in the HO case, and for $\La \geq 16$ in the DW and AHO cases.} \label{fig:energy_spectrum_all_1}
\end{figure}

Without further ado, \fig{fig:energy_spectrum_all_1} presents the $m=g=\mu=1$ energy spectrum for each of the three superpotentials, as \La increases from 2 to 32.  The six lowest-lying energies are shown.
We see that the ground state energy rapidly converges for $\La > 2$, and also quickly achieving the correct pairing structure (degenerate with the first excited state for the DW superpotential but not for the HO and AHO superpotentials).
For excited states, the HO superpotential exhibits fast convergence for $\La > 4$, while the DW and AHO cases require $\La \geq 16$. These results confirm the expected behavior that increasing the number of bosonic modes reduces truncation effects, especially evident in the excited-state spectrum.

\begin{figure}[htbp]
    \centering
    \begin{subfigure}{0.48\textwidth}
        \centering
        \includegraphics[width=\linewidth]{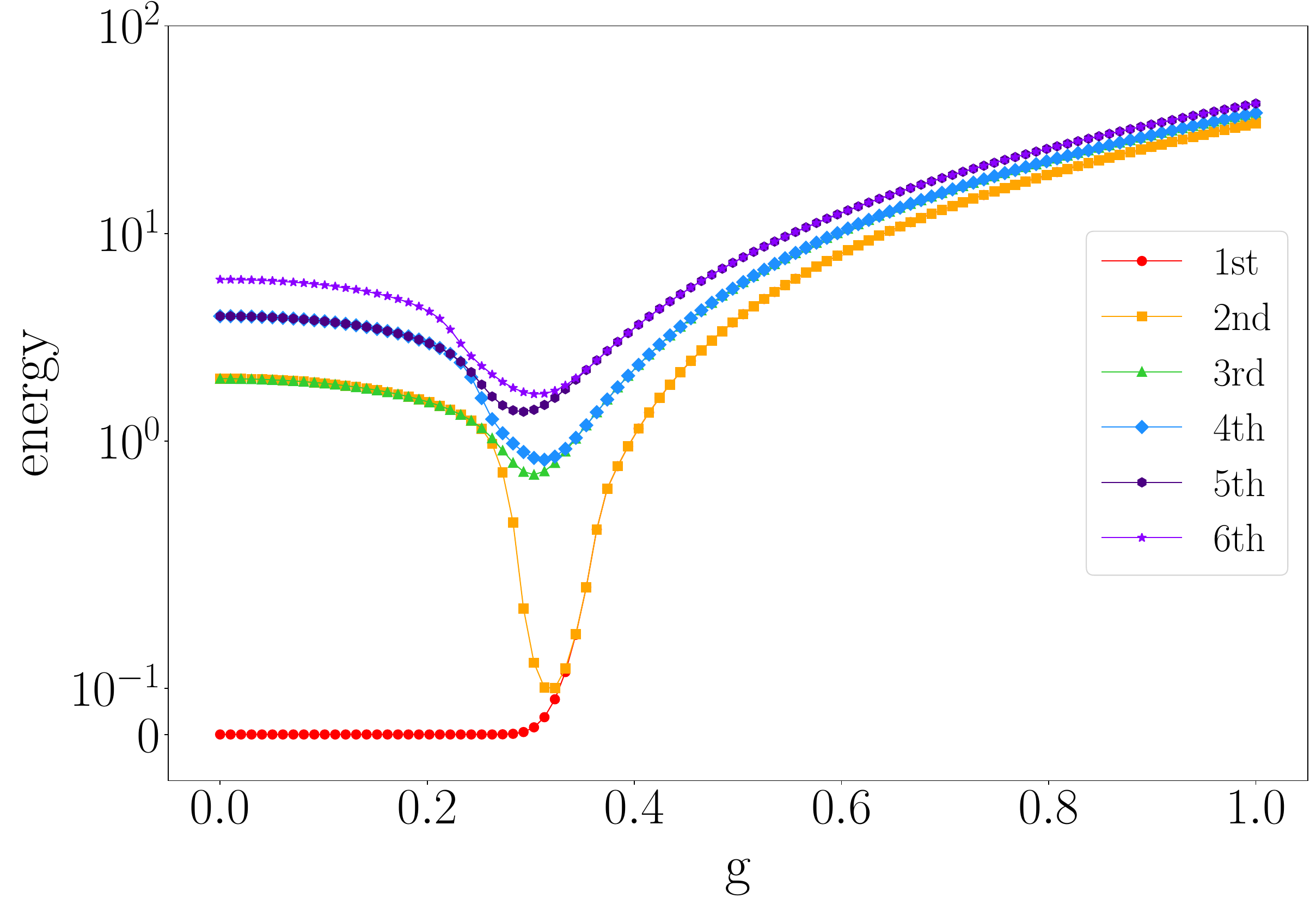}
        {DW, $\La = 32$}
    \end{subfigure}
    \hfill
    \begin{subfigure}{0.48\textwidth}
        \centering
    \includegraphics[width=\linewidth]{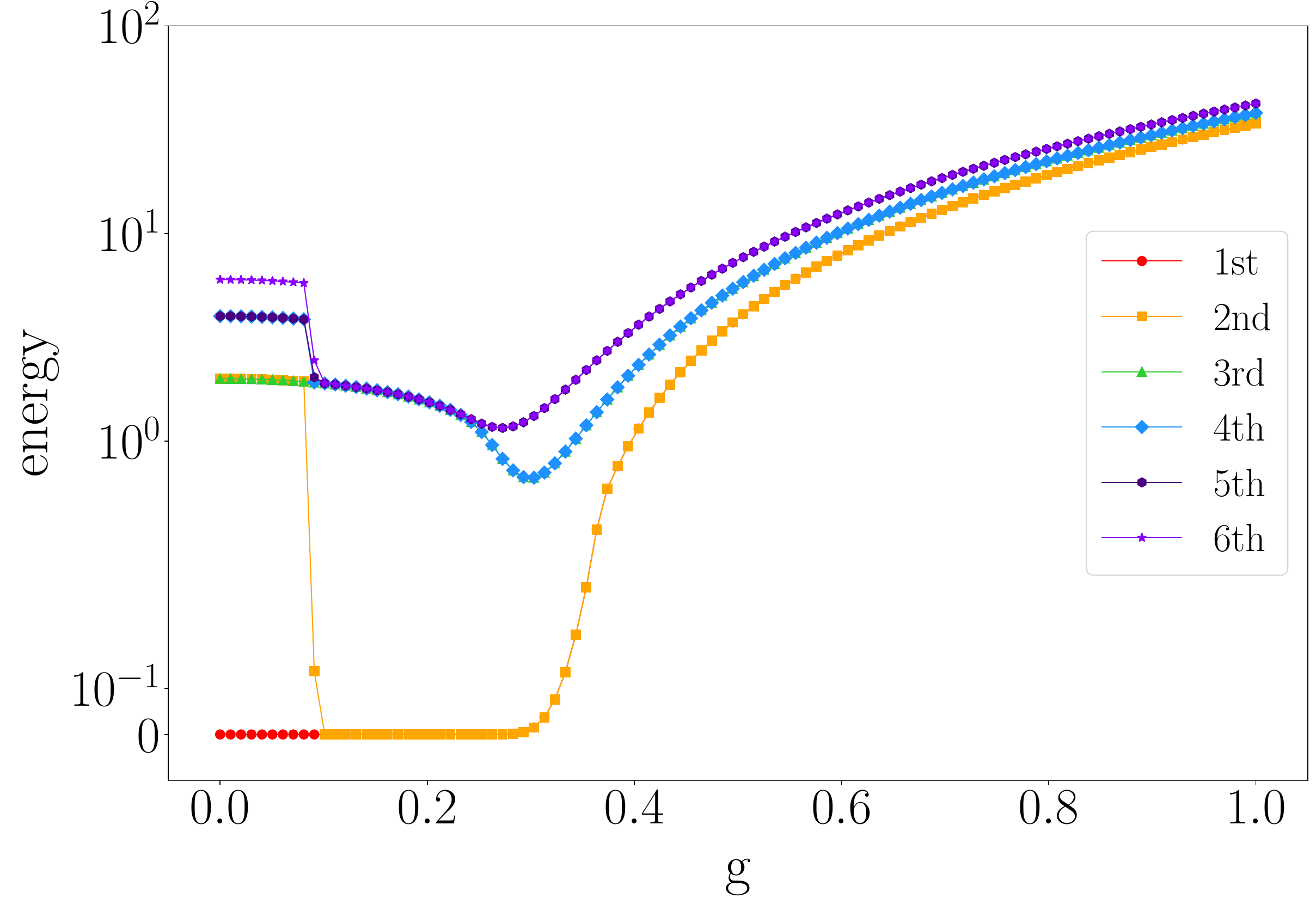}
   {DW, $\La = 512$}
    \end{subfigure}
    \caption{The low-lying energy spectrum from exact diagonalization, consisting of the first six energy levels, for the case of the DW superpotential with fixed $m= 2$ and $\mu=3$. Each energy level is represented by a marker connected with a thin line to guide the eye.  We vary $g$ from 0 to 1, and compare two values of \La to highlight the effects of truncating the bosonic degree of freedom. On the left, the energy spectrum is displayed for $\La = 32$, and on the right for $\La = 512$. These two plots confirm that the DW case exhibits spontaneous supersymmetry breaking for all $g > 0$ when $\La \to \infty$.} \label{fig:DW_energy_spectrum}
\end{figure}

Now let us generalize to parameters $(m, g, \mu)$ different from unity.
In the HO case $m$ is the only parameter, and varying its value just shifts the energies without altering the pairing structure.
In the DW case, all three parameters are present.
As $g \to 0$, the superpotential approaches the HO and the spectrum should therefore be supersymmetric, whereas spontaneous supersymmetry breaking is expected for the larger $g \gtrsim 1$ considered in the body of the paper.
This is confirmed in \fig{fig:DW_energy_spectrum} for fixed $m= 2$ and $\mu=3$.
This figure also shows that the value of $g$ around which the pairing structure changes decreases from $g \approx 0.32$ for $\La = 32$ to $g \approx 0.086$ for $\La = 512$.
We can expect that upon removing the truncation by extrapolating $\La \to \infty$, the DW case exhibits spontaneous supersymmetry breaking for any positive $g > 0$.

\begin{figure}[htpb]
    \centering
    \begin{subfigure}{0.48\textwidth}
        \centering
        \includegraphics[width=\linewidth]{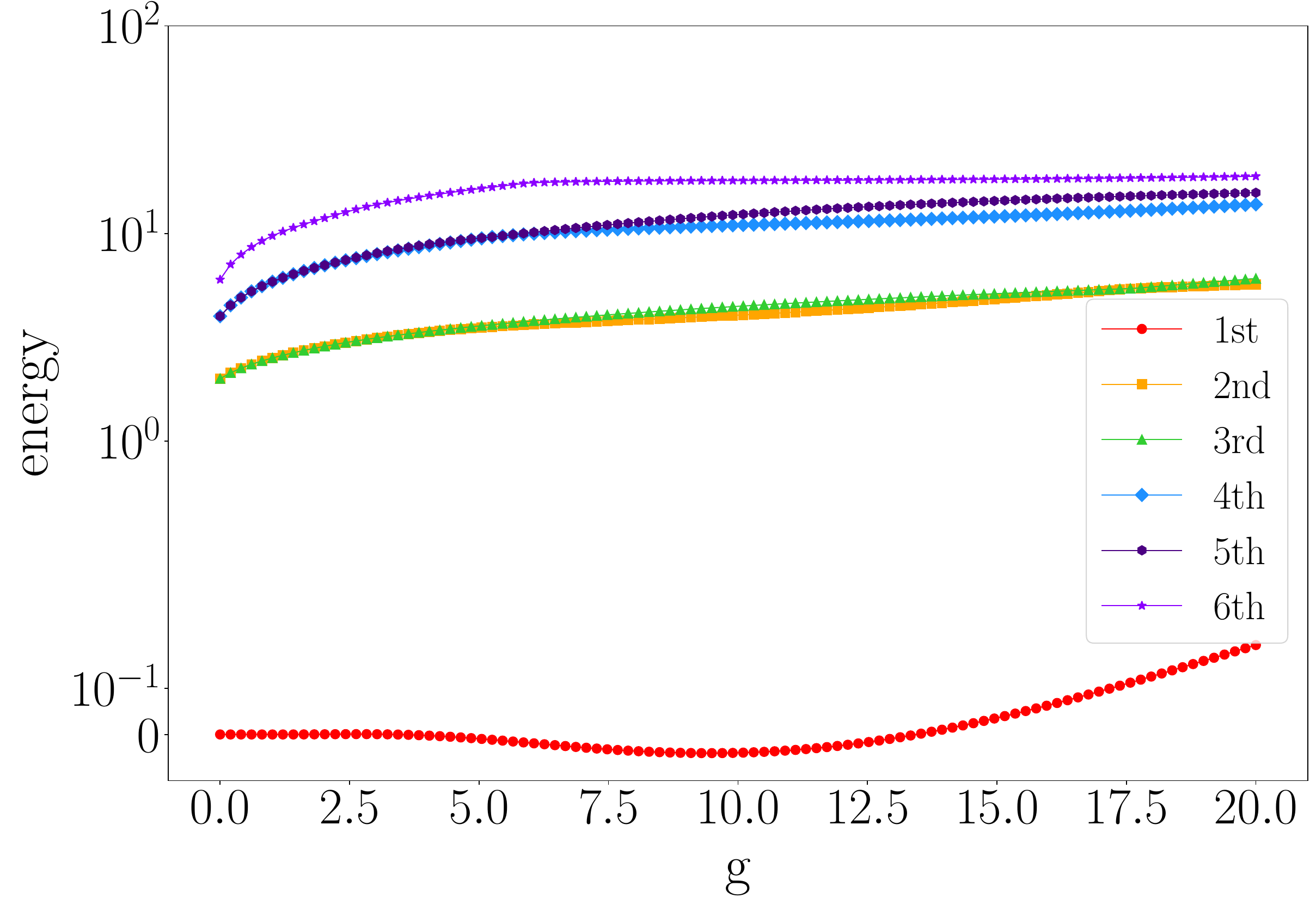}
        {AHO, $\La = 16$}
    \end{subfigure}
    \hfill
    \begin{subfigure}{0.48\textwidth}
        \centering
        \includegraphics[width=\linewidth]{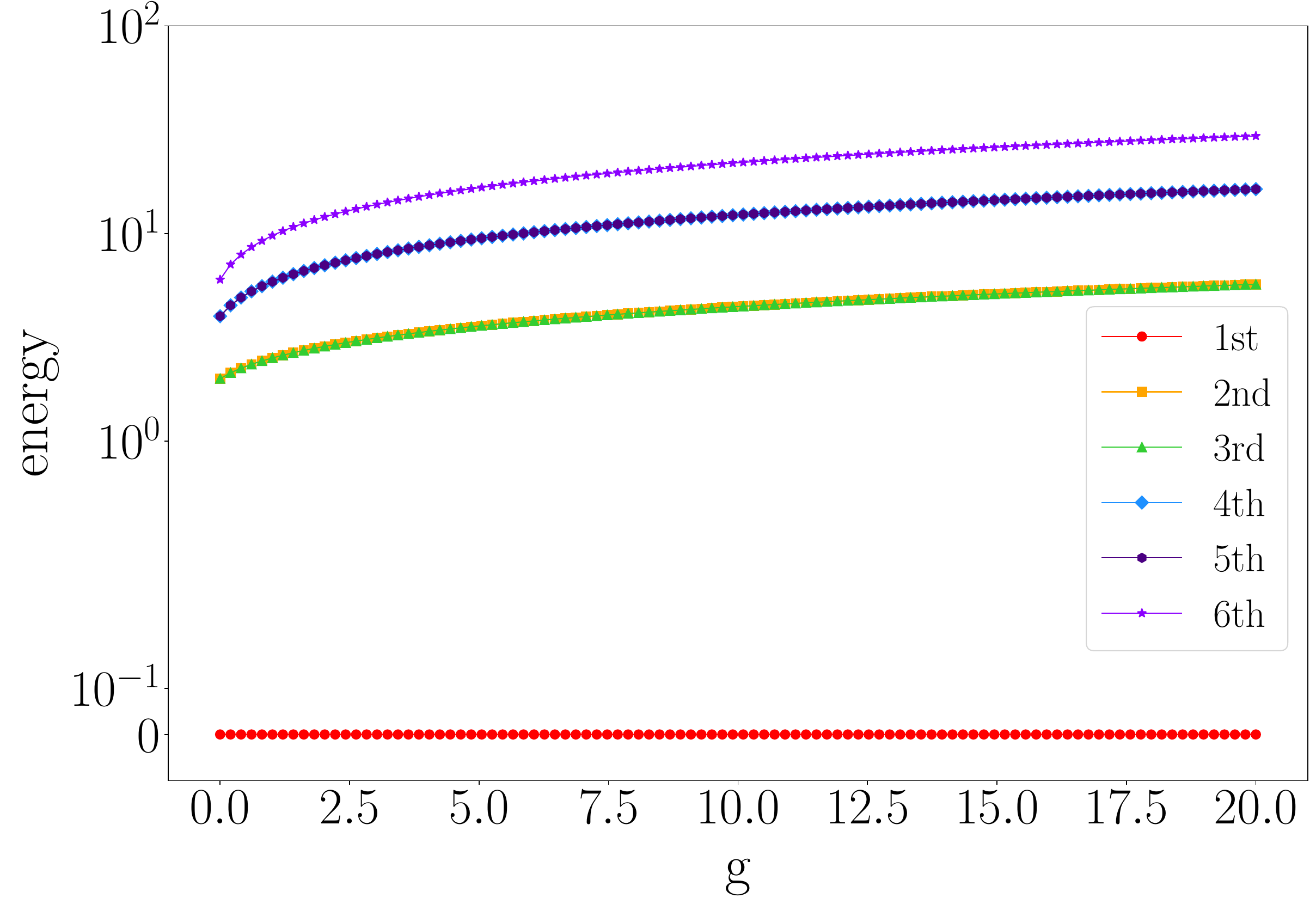}
   {AHO, $\La = 512$}
    \end{subfigure}
    \caption{The low-lying energy spectrum from exact diagonalization, as in \fig{fig:DW_energy_spectrum}, but now for the case of the AHO superpotential with fixed $m = 2$. We now vary $g$ from 0 to 20, and compare $\La = 16$ and 512 to highlight the effect of bosonic mode truncation. On the left, the energy spectrum for $\La = 16$ shows sign of truncation artifacts in the growing ground state energy and the disruption of the exited state paring structure as $g$ is increased. Those effects are suppressed as \La increases, as shown on the right for $\La = 512$.} \label{fig:AHO_energy_spectrum}
\end{figure}

In the AHO case, only the two parameters $m$ and $g$ are present.
Fixing $m = 2$, \fig{fig:AHO_energy_spectrum} shows the $\La = 16$ ground-state energy rising from zero to a non-zero value as $g$ increases, and the characteristic excited-state pairing structure becomes is slightly disrupted (though it remains obvious that supersymmetry is not spontaneously broken). Although the energy spectrum improves significantly already for $\La = 32$, \fig{fig:AHO_energy_spectrum} compares the $\La = 16$ and $\La = 512$ energy spectra, emphasizing the similarity for small $g \lesssim 5$.

As shown in Figs.~\ref{fig:DW_energy_spectrum} and \ref{fig:AHO_energy_spectrum}, truncation artifacts in the first few energy levels can make it challenging to determine whether supersymmetry is preserved or spontaneously broken, and we expect the situation to get worse when using noisy NISQ hardware.
The ratio $R$ introduced in \eq{eq:vqd-ratio} can help to address this issue more quantitatively.
Recall that $R$ vanishes when preserved supersymmetry makes the first two excited states degenerate, $E_1 \approx E_2$, while $R \approx 1$ when the energy of the first excited state approaches the ground-state energy, $E_0 \approx E_1$, signaling spontaneous supersymmetry breaking.
We provide results for $R$ for each superpotential with \La up to 4096 in Tables~\ref{table:HO_ratio} (HO), \ref{table:DW_ratio} (DW) and~\ref{table:AHO_ratio} (AHO).

\begin{table}[htpb]
\centering
\begin{adjustbox}{width=\textwidth}
\centering\begin{tabular}{c|c|c|c|c|c}
\hline
\hline
\La & Ground State & 1st Excited & 2nd Excited & $R$ &  Supersymmetry \\
\hline
2 & $0\mathrm{e}{+00}$ & $0\mathrm{e}{+00}$ & $1\mathrm{e}{+00}$ & $1\mathrm{e}{+00}$ & \text{\sffamily X} \\
4 & $0\mathrm{e}{+00}$ & $1\mathrm{e}{+00}$ & $1\mathrm{e}{+00}$ & $0\mathrm{e}{+00}$ & \checkmark \\
16 & $0\mathrm{e}{+00}$ & $1\mathrm{e}{+00}$ & $1\mathrm{e}{+00}$ & $2.22044605\mathrm{e}{-16}$ & \checkmark \\
32 & $0\mathrm{e}{+00}$ & $1\mathrm{e}{+00}$ & $1\mathrm{e}{+00}$ & $2.22044605\mathrm{e}{-16}$ & \checkmark \\
64 & $0\mathrm{e}{+00}$ & $1\mathrm{e}{+00}$ & $1\mathrm{e}{+00}$ & $2.22044605\mathrm{e}{-16}$ & \checkmark \\
128 & $0\mathrm{e}{+00}$ & $1\mathrm{e}{+00}$ & $1\mathrm{e}{+00}$ & $2.22044605\mathrm{e}{-16}$ &  \checkmark\\
256 & $0\mathrm{e}{+00}$ & $1\mathrm{e}{+00}$ & $1\mathrm{e}{+00}$ & $6.43929354\mathrm{e}{-15}$ & \checkmark \\
512 & $0\mathrm{e}{+00}$ & $1\mathrm{e}{+00}$ & $1\mathrm{e}{+00}$ & $6.43929354\mathrm{e}{-15}$ &  \checkmark\\
1024 & $0\mathrm{e}{+00}$ & $1\mathrm{e}{+00}$ & $1\mathrm{e}{+00}$ & $6.43929354\mathrm{e}{-15}$ & \checkmark \\
2048 & $0\mathrm{e}{+00}$ & $1\mathrm{e}{+00}$ & $1\mathrm{e}{+00}$ & $6.43929354\mathrm{e}{-15}$ & \checkmark \\
4096 & $8.73509654\mathrm{e}{-12}$ & $1\mathrm{e}{+00}$ & $1\mathrm{e}{+00}$ & $7.98183741\mathrm{e}{-12}$ & \checkmark \\
\hline
\hline
\end{tabular}
\end{adjustbox}
\caption{The first three energy levels, and the numerical value of the ratio $R$ in \eq{eq:vqd-ratio}, for SQM with the Harmonic Oscillator superpotential and \La bosonic modes.  The final column indicates whether $R$ implies spontaneous supersymmetry breaking (\text{\sffamily X}) or preserved supersymmetry (\checkmark).}\label{table:HO_ratio}
\end{table}

\begin{table}[htbp]
\centering
\begin{adjustbox}{width=\textwidth}
\centering\begin{tabular}{c|c|c|c|c|c}
\hline
\hline
\La & Ground State & 1st Excited & 2nd Excited & $R$ &  Supersymmetry \\
\hline
2 & $3.57233047\mathrm{e}{-01}$ & $7.71446609\mathrm{e}{-01}$ & $1.47855339\mathrm{e}{+00}$ & $6.30601937\mathrm{e}{-01}$ & \text{\sffamily ?} \\
4 & $9.06559871\mathrm{e}{-01}$ & $9.50633533\mathrm{e}{-01}$ & $1.69566635\mathrm{e}{+00}$ & $9.44147383\mathrm{e}{-01}$ & \text{\sffamily X} \\
8 & $8.84580444\mathrm{e}{-01}$ & $8.87725463\mathrm{e}{-01}$ & $2.69387284\mathrm{e}{+00}$ & $9.98261741\mathrm{e}{-01}$ &  \text{\sffamily X}\\
16 & $8.91599362\mathrm{e}{-01}$ & $8.91640946\mathrm{e}{-01}$ & $2.73412225\mathrm{e}{+00}$ & $9.99977431\mathrm{e}{-01}$ & \text{\sffamily X}\\
32 & $8.91632371\mathrm{e}{-01}$ & $8.91632381\mathrm{e}{-01}$ & $2.7340106\mathrm{e}{+00}$ & $9.99999994\mathrm{e}{-01}$ & \text{\sffamily X}\\
64 & $8.9163238\mathrm{e}{-01}$ & $8.9163238\mathrm{e}{-01}$ & $2.73401076\mathrm{e}{+00}$ & $1\mathrm{e}{+00}$ & \text{\sffamily X}\\
128 & $8.9163238\mathrm{e}{-01}$ & $8.9163238\mathrm{e}{-01}$ & $2.73401076\mathrm{e}{+00}$ & $1\mathrm{e}{+00}$ & \text{\sffamily X}\\
256 & $8.9163238\mathrm{e}{-01}$ & $8.9163238\mathrm{e}{-01}$ & $2.73401076\mathrm{e}{+00}$ & $1\mathrm{e}{+00}$ & \text{\sffamily X}\\
512 & $8.9163238\mathrm{e}{-01}$ & $8.9163238\mathrm{e}{-01}$ & $2.73401076\mathrm{e}{+00}$ & $1\mathrm{e}{+00}$ & \text{\sffamily X}\\
1024 & $8.9163238\mathrm{e}{-01}$ & $8.9163238\mathrm{e}{-01}$ & $2.73401076\mathrm{e}{+00}$ & $1\mathrm{e}{+00}$ & \text{\sffamily X}\\
2048 & $8.9163238\mathrm{e}{-01}$ & $8.9163238\mathrm{e}{-01}$ & $2.73401076\mathrm{e}{+00}$ & $1\mathrm{e}{+00}$ & \text{\sffamily X}\\
4096 & $8.91632371\mathrm{e}{-01}$ & $8.91632409\mathrm{e}{-01}$ & $2.73401077\mathrm{e}{+00}$ & $9.9999998\mathrm{e}{-01}$ & \text{\sffamily X} \\
\hline
\hline
\end{tabular}
\end{adjustbox}
\caption{Similar to \tab{table:HO_ratio}, but for the Double Well superpotential.}\label{table:DW_ratio}
\end{table}

\begin{table}[htpb]
\centering
\begin{adjustbox}{width=\textwidth}
\centering\begin{tabular}{c|c|c|c|c|c}
\hline
\hline
\La & Ground State & 1st Excited & 2nd Excited & $R$ &  Supersymmetry \\
\hline
2 & $-4.375\mathrm{e}{-01}$ & $-4.375\mathrm{e}{-01}$ & $2.0625\mathrm{e}{+00}$ & $1\mathrm{e}{+00}$ & \text{\sffamily X}\\
4 & $-1.64785261\mathrm{e}{-01}$ & $6.73310098\mathrm{e}{-01}$ & $1.66794264\mathrm{e}{+00}$ & $5.42706061\mathrm{e}{-01}$ & \text{\sffamily ?}\\
8 & $3.201011\mathrm{e}{-02}$ & $1.68015552\mathrm{e}{+00}$ & $1.83352558\mathrm{e}{+00}$ & $8.51339075\mathrm{e}{-02}$ & \checkmark\\
16 & $-1.16697568\mathrm{e}{-03}$ & $1.6774942\mathrm{e}{+00}$ & $1.68638125\mathrm{e}{+00}$ & $5.26625106\mathrm{e}{-03}$ & \checkmark\\
32 & $6.1822377\mathrm{e}{-06}$ & $1.68649858\mathrm{e}{+00}$ & $1.68655905\mathrm{e}{+00}$ & $3.58545635\mathrm{e}{-05}$ & \checkmark\\
64 & $-1.94552515\mathrm{e}{-08}$ & $1.68650037\mathrm{e}{+00}$ & $1.68650053\mathrm{e}{+00}$ & $9.73093462\mathrm{e}{-08}$ & \checkmark\\
128 & $4.56724565\mathrm{e}{-11}$ & $1.68650053\mathrm{e}{+00}$ & $1.68650053\mathrm{e}{+00}$ & $1.3456307\mathrm{e}{-11}$ & \checkmark\\
256 & $-2.22142414\mathrm{e}{-10}$ & $1.68650053\mathrm{e}{+00}$ & $1.68650053\mathrm{e}{+00}$ & $8.67340318\mathrm{e}{-11}$ & \checkmark\\
512 & $-2.55886884\mathrm{e}{-10}$ & $1.68650053\mathrm{e}{+00}$ & $1.68650053\mathrm{e}{+00}$ & $1.642518\mathrm{e}{-09}$ & \checkmark \\
1024 & $4.18257223\mathrm{e}{-10}$ & $1.68650052\mathrm{e}{+00}$ & $1.68650054\mathrm{e}{+00}$ & $1.1430257\mathrm{e}{-08}$ & \checkmark \\
2048 & $-2.96934283\mathrm{e}{-08}$ & $1.68650056\mathrm{e}{+00}$ & $1.68650059\mathrm{e}{+00}$ & $1.95532552\mathrm{e}{-08}$ & \checkmark \\
4096 & $-5.23873122\mathrm{e}{-07}$ & $1.68650038\mathrm{e}{+00}$ & $1.68650051\mathrm{e}{+00}$ & $7.50023801\mathrm{e}{-08}$ & \checkmark \\
\hline
\hline
\end{tabular}
\end{adjustbox}
\caption{Similar to the \tab{table:HO_ratio}, but for the Anharmonic Oscillator superpotential.}\label{table:AHO_ratio}
\end{table}

\section{Block structure}\label{app:block-structure}
In this appendix we take a closer look at the structure of the Hamiltonian in \eq{eq:H_SQM}, explicitly showing the tensor products between operators acting on different subspaces:
\begin{equation}\label{eq:H_block}
  H =I_f \otimes \frac{1}{2}\left( \phat^2 + [W'(\qhat)]^2 \right) - \frac{1}{2} \left( \left[\bhatdag, \bhat\right] \otimes W''(\qhat) \right).
\end{equation}
Here the identity operator in the fermionic subspace, $I_f$, has size $2 \times 2$.
To reveal the block structure, it is sufficient to expand
\begin{align*}
  I_f & = \ket{0} \bra{0} + \ket{1} \bra{1} &
  \left[\bhatdag, \bhat\right] & = \ket{1} \bra{1} - \ket{0} \bra{0}
\end{align*}
in the fermionic subspace. The commutator follows from the explicit expressions for \bhat and \bhatdag in \eq{eq:b_bdag_matrix_r}.
This leads to
\begin{align}\label{eq:H_block_final}
  H &= \left( \ket{0} \bra{0} + \ket{1} \bra{1}  \right) \otimes \frac{1}{2}\left( \phat^2 + [W'(\qhat)]^2 \right) - \frac{1}{2} \left( \left( \ket{1} \bra{1} - \ket{0} \bra{0}  \right) \otimes W''(\qhat) \right) \\
  &= \ket{0} \bra{0} \otimes \frac{1}{2}\left( \phat^2 + [W'(\qhat)]^2 + W''(\qhat) \right) + \ket{1} \bra{1} \otimes \frac{1}{2}\left( \phat^2 + [W'(\qhat)]^2 -  W''(\qhat) \right) \nonumber \\
  &=\left( \begin{matrix} 1&0\\0&0 \end{matrix} \right)  \otimes \frac{1}{2}\left( \phat^2 + [W'(\qhat)]^2 + W''(\qhat) \right) + \left( \begin{matrix} 0&0\\0&1 \end{matrix} \right) \otimes \frac{1}{2}\left( \phat^2 + [W'(\qhat)]^2 - W''(\qhat) \right)\nonumber \, ,
\end{align}
where the last expression clearly shows the Hamiltonian block structure:
\begin{equation}\label{eq:H_block_structure}
\small
H=\left(
\begin{array}{ccc|ccc}
& & & & & \\
& \frac{1}{2}\left( \phat^2 + [W'(\qhat)]^2 + W''(\qhat) \right) & & \text{\Large 0}_{\La \times \La} & \\
& & & & & \\
\hline
& & & & & \\
& \text{\Large 0}_{\La \times \La} & & \quad \quad \frac{1}{2}\left( \phat^2 + [W'(\qhat)]^2 - W''(\qhat) \right) & \\
& & & & & \\
\end{array}
\right)
\begin{matrix}
    \ket{0}\ket{0} \\ \dots \\ \ket{0}\ket{\La - 1} \\
    \, \\
    \ket{1}\ket{0} \\ \dots \\ \ket{1}\ket{\La - 1} \\
\end{matrix}
\end{equation}

Finally, the action of the Hamiltonian on a generic state $\ket{\Psi} = \ket{f} \ket{b}$ is given by
\begin{align} \label{eq:H_on_state}
  H \ket{\Psi} = H \ket{f} \ket{b} & = \left( \begin{matrix} 1&0\\0&0 \end{matrix} \right) \ket{f} \delta_{0,f} \otimes \frac{1}{2}\left( \phat^2 + [W'(\qhat)]^2 + W''(\qhat) \right) \ket{b} \\
  &+ \left( \begin{matrix} 0&0\\0&1 \end{matrix} \right) \ket{f} \delta_{1,f} \otimes \frac{1}{2}\left( \phat^2 + [W'(\qhat)]^2 - W''(\qhat) \right) \ket{b}. \nonumber
\end{align}
Therefore the top-right block contains $\ket{0}\ket{b}$ elements, while the bottom-left block the $\ket{1} \ket{b}$ elements as shown in \eq{eq:H_block_structure}.
Furthermore, this implies that the expectation value $\bra{\Psi'} H \ket{\Psi} = \bra{f'} \bra{b'} H \ket{f} \ket{b}$ is zero unless the fermionic states $\ket{f}$ and $\ket{f'}$ coincide.

Due to the block structure, the ground state is associated with one of the two blocks.
Specifically, for the HO and AHO superpotentials with a sufficiently large number of bosonic modes, $\La \geq 8$, the ground state belongs to the lower block, $\ket{1}\ket{b}$.
In the DW case, it belongs to the upper block $\ket{0}\ket{b}$.

From a practical perspective, knowing that the Hamiltonian of the model consists of two decoupled blocks means that, in the search for the ground state, only one block needs to be encoded on quantum hardware. This results in a reduction of the resources required to analyze the system, as shown by \tab{tab:block_H_pauli_strings_energybasis}.  Compared to \tab{tab:full_H_pauli_strings_energybasis} in Section~\ref{sec:fock_basis_digitization}, the number of qubits corresponding to a given \La is reduced by one, while the number of Pauli strings decreases by approximately $30\%$ for the DW and AHO superpotentials. Furthermore, the block structure was instrumental in enhancing the efficiency and convergence of the adaptive methods presented in Section~\ref{section:Adaptive-VQE}, as it facilitated the selection of an appropriate initial state to construct the ground-state ansatz within the correct block for each superpotential.

\begin{table}[htpb]
\centering
\begin{adjustbox}{width=\textwidth}
\begin{tabular}{cccccc}
\hline
\hline
\La & $H$ size & $n$ Qubits & Harmonic Oscillator & Double Well & Anharmonic Oscillator\\ \hline
2 & $2\times 2$ & 1 & 0 & 2 & 1\\
4 & $4\times 4$ & 2 & 3 & 9 & 5\\
8 & $8\times 8$ & 3 & 7 & 35 & 19\\
16 & $16\times 16$ & 4 & 15 & 103 & 63\\
32 & $32\times 32$ & 5 & 31 & 271 & 175\\
64 & $64\times 64$ & 6 & 63 & 661 & 447\\
128 & $128\times 128$ & 7 & 127 & 1541 & 1079\\
256 & $256\times 256$ & 8 & 255 & 3425 & 2499\\
512 & $512\times 512$ & 9 & 511 & 7569 & 5459\\
1024 & $1024\times 1024$ &10 & 1023 & 16081 & 11551\\
2048 & $2048 \times 2048$ &11 & 2047 & 33743 & 24039\\
\hline
\hline
\end{tabular}
\end{adjustbox}
\caption{A summary of the quantum resources needed to encode the block of the Hamiltonian that contains the ground state, for \La bosonic modes with $m = g = \mu = 1$. The number of Pauli strings required for each superpotential is reduced compared to \tab{tab:full_H_pauli_strings_energybasis}.}\label{tab:block_H_pauli_strings_energybasis}
\end{table}

\addcontentsline{toc}{section}{\refname}
\printbibliography
\end{document}